\documentclass[aps,pre,onecolumn,superscriptaddress,showpacs]{revtex4-2}
\usepackage{graphicx}
\usepackage{array}
\usepackage{amsfonts}
\usepackage{amssymb,amsmath,multirow,rotate}
\DeclareMathOperator*{\argmin}{argmin}
\usepackage{mathrsfs}
\usepackage{booktabs}
\usepackage{threeparttable}
\usepackage{multirow}
\usepackage{epsfig}
\usepackage{threeparttable}
\usepackage{chngpage}
\usepackage{latexsym}
\usepackage{dcolumn}
\usepackage{graphics,graphicx,bm,fleqn,epic,eepic,float}
\usepackage{verbatim}
\usepackage{subfigure}
\usepackage{color}
\usepackage{xr}
\usepackage{float}
\usepackage{listings} % For including code in appendix
\usepackage{placeins} % For putting plots in the desired location
\usepackage{tikz}
\usepackage[pdfborder={0 0 0}]{hyperref}
\usepackage{xurl}
\usepackage{xcolor}
\usepackage{soul}

\definecolor{red}{rgb}{1,0,0}

\definecolor{blue}{rgb}{0,0,1}

\usepackage{caption}
\begin{document}

\title{Validity of Markovian modeling for transient memory-dependent epidemic dynamics}

\author{Mi Feng}
\affiliation{Department of Physics, Hong Kong Baptist University, Kowloon Tong, Hong Kong SAR 999077, China}
\affiliation{Centre for Nonlinear Studies and Beijing-Hong Kong-Singapore Joint Centre for Nonlinear and Complex Systems (Hong Kong), Hong Kong Baptist University, Kowloon Tong, Hong Kong SAR 999077, China}

\author{Liang Tian} \email{liangtian@hkbu.edu.hk}
\affiliation{Department of Physics, Hong Kong Baptist University, Kowloon Tong, Hong Kong SAR 999077, China}
\affiliation{Institute of Computational and Theoretical Studies, Hong Kong Baptist University, Kowloon, Hong Kong SAR 999077, China}

\author{Ying-Cheng Lai}
\affiliation{School of Electrical, Computer and Energy Engineering, Arizona State University, Tempe, AZ 85287, USA}
\affiliation{Department of Physics, Arizona State University,
Tempe, Arizona 85287, USA}

\author{Changsong Zhou} \email{cszhou@hkbu.edu.hk}
\affiliation{Department of Physics, Hong Kong Baptist University, Kowloon Tong, Hong Kong SAR 999077, China}
\affiliation{Centre for Nonlinear Studies and Beijing-Hong Kong-Singapore Joint Centre for Nonlinear and Complex Systems (Hong Kong), Hong Kong Baptist University, Kowloon Tong, Hong Kong SAR 999077, China}
\affiliation{Institute of Computational and Theoretical Studies, Hong Kong Baptist University, Kowloon, Hong Kong SAR 999077, China}

\begin{abstract}

The initial transient phase of an emerging epidemic is of critical importance for data-driven model building, model-based prediction of the epidemic trend, and articulation of control/prevention strategies. In principle, quantitative models for real-world epidemics need to be memory-dependent or non-Markovian, but this presents difficulties for data collection, parameter estimation, computation and analyses. In contrast, the difficulties do not arise in the traditional Markovian models. To uncover the conditions under which Markovian and non-Markovian models are equivalent for transient epidemic dynamics is outstanding and of significant current interest. We develop a comprehensive computational and analytic framework to establish that the transient-state equivalence holds when the average generation time matches and average removal time, resulting in minimal Markovian estimation errors in the basic reproduction number, epidemic forecasting, and evaluation of control strategy. Strikingly, the errors depend on the generation-to-removal time ratio but not on the specific values and distributions of these times, and this universality will further facilitate estimation rectification. Overall, our study provides a general criterion for modeling memory-dependent processes using the Markovian frameworks.

\end{abstract}

\maketitle

\section*{Introduction}

When an epidemic emerges, the initial transient phase of the disease spreading 
dynamics before a steady state is reached is of paramount importance, for two 
reasons~\cite{dong2020interactive,mishra2021one,ferretti2020quantifying,giordano2021modeling,
jentsch2021prioritising,bubar2021model,buckner2021dynamic,goldstein2021vaccinating,viana2021controlling,matrajt2021vaccine,matrajt2021optimizing, zhao2020preliminary,billah2020reproductive,liu2022effective}. First, key indicators or parameters characterizing the underlying 
dynamical process and critical for prediction and articulation of control
strategies, such as the generation time, the serial intervals and the basic 
reproduction number, are required to be estimated when the dynamics have not reached a 
steady state. Second, it is during the transient phase control and mitigation 
strategies can be effectively applied for preventing a large scale outbreak.
Prediction and control depend, of course, on a quantitative model of the 
epidemic process, which can be constructed based on the key parameters estimated 
from data collected during the transient phase. In principle, since the dynamical 
processes underlying real-world epidemics are generally memory-dependent in the
sense that the state evolution depends on the history, a rigorous modeling 
framework needs to be of the non-Markovian type, but this presents great 
challenges in terms of data collection, parameter estimation, computation and 
analyses~\cite{metcalf2017opportunities, riley2007large, levin1997mathematical}. 
The difficulties can be significantly alleviated by adopting the 
traditional memoryless, Markovian framework ~\cite{pastor2015epidemic,
lambiotte2013burstiness,lin2020non,cator2013susceptible,
boguna2014simulating,van2013non,starnini2017equivalence,feng2019equivalence,
min2013suppression,karrer2010message,wang2003epidemic,
chakrabarti2008epidemic}. An outstanding question is, under 
what conditions will an approximate equivalence between non-Markovian and 
Markovian dynamics hold during the {\em transient phase} of the epidemic? 
Additionally, another important issue remains, how are the errors of Markovian estimation determined? The 
purpose of this paper is to a comprehensive answer to these questions.

The COVID-19 pandemic has highlighted the need and importance of understanding
disease spreading and transmission to accurately predict, control and manage 
future outbreaks through non-pharmacological interventions and vaccine 
allocation strategies~\cite{dong2020interactive,mishra2021one,ferretti2020quantifying,giordano2021modeling,jentsch2021prioritising,bubar2021model,buckner2021dynamic,goldstein2021vaccinating,viana2021controlling,matrajt2021vaccine,matrajt2021optimizing}. To accomplish these goals, accurate mathematical modeling of the disease 
spreading dynamics is key. In a general population, epidemic transmission occurs 
via some kind of point process, where individuals become infected at different 
points in time. It has been known that point processes in the real world are 
typically non-Markovian with a memory effect in which the distribution of the 
interevent times is not exponential~\cite{ferretti2020quantifying,keeling1997disease,barabasi2005origin,gonzalez2008understanding,simini2012universal,pappalardo2015returners,yan2017universal,bratsun2005delay,scalas2006waiting,vazquez2007impact,min2011spreading,min2013suppression,moghadas2020implications,lauer2020incubation,gatto2020spread,li2020substantial,buitrago2020occurrence,world2003consensus,chowell2004model,eichner2003transmission}. For example, the interevent time distribution arising from the virus transmission with COVID-19 is not of the memoryless exponential 
type but typically exhibits memory-dependent features characterized by the Weibull 
distribution~\cite{ferretti2020quantifying}. Strictly speaking, from a modeling 
perspective, disease spreading should be described by a non-Markovian process. 
A non-Markovian approach takes into account historical memory of disease 
progression, mathematically resulting in a complex set of integro-differential 
equations in the form of convolution.

There are significant difficulties with non-Markovian modeling of memory-dependent 
disease spreading. The foremost is data availability. In particular, while 
standard epidemic spreading models are available, the model parameters need to be 
estimated through data. A non-Markovian model often requires detailed and granular
data that can be difficult to get, especially during the early stage of the 
epidemic where accurate modeling is most needed~\cite{metcalf2017opportunities}. 
From a theoretical point of view, it is desired to obtain certain closed-form 
solutions for key quantities such as the onset and size of the epidemic outbreak, 
but this is generally impossible for non-Markovian models~\cite{pastor2015epidemic,grassly2008mathematical}. Computationally, accommodating memory effects in 
principle makes the underlying dynamical system infinitely dimensional, 
practically requiring solving an unusually large number of dynamical variables 
through a large number of complex integro-differential equations~\cite{riley2007large,levin1997mathematical}. In contrast, in an idealized Markovian point process, 
events occur at a fixed rate, leading to an exponential distribution for the 
interevent time intervals and consequently a memoryless process. If the spreading 
dynamics were of the Markovian type, the aforementioned difficulties associated with
non-Markovian dynamics no longer exist. In particular, a Markovian spreading 
process can be described by a small number of ordinary differential equations 
with a few parameters that can be estimated even from sparse data, and the 
numerical simulations can be carried out in a computationally extremely efficient 
manner~\cite{pastor2015epidemic,lambiotte2013burstiness,lin2020non,cator2013susceptible,boguna2014simulating,van2013non,starnini2017equivalence,feng2019equivalence,min2013suppression,karrer2010message,wang2003epidemic,chakrabarti2008epidemic}. For these reasons, many recent studies of the COVID-19 pandemic 
assumed Markovian behaviors to avoid or ``escape'' from the difficulties 
associated with non-Markovian modeling~\cite{giordano2021modeling,jentsch2021prioritising,bubar2021model,buckner2021dynamic,goldstein2021vaccinating,viana2021controlling,matrajt2021vaccine,matrajt2021optimizing}. The issue is whether such a 
simplified approach can be justified. Addressing this issue requires a 
comprehensive understanding of the extent to which the Markovian approach 
represents a good approximation to model non-Markovian type of memory-dependent 
spreading dynamics, and specifically of the conditions under which the Markovian theory can produce accurate results that match those from the non-Markovian model.

There were previous studies of the so-called steady-state equivalence between 
Markovian and non-Markovian modeling for epidemic spreading. In particular, when 
the system has reached a steady state, such an equivalence can be established 
through a modified definition of the effective infection rate~\cite{starnini2017equivalence,feng2019equivalence,cator2013susceptible}. From a realistic point of 
view, the equivalence limited only to steady states may not be critical as the 
transient phase of the spreading process before any steady state is reached is 
more relevant and significant. For example, when an epidemic occurs, it is of 
fundamental interest to estimate the key indicators such as the generation time
(the time interval between the infections of the infector and infectee in a 
transmission chain), the serial interval (the time from illness onset in the 
primary case to illness onset in the secondary case), and the basic reproduction 
number (the average number of secondary transmissions from one infected person),
but they are often needed to be estimated when the dynamics have not reached a steady 
state~\cite{zhao2020preliminary,billah2020reproductive,liu2022effective,ferretti2020quantifying,
giordano2021modeling,jentsch2021prioritising,bubar2021model,buckner2021dynamic,
goldstein2021vaccinating,viana2021controlling,matrajt2021vaccine,matrajt2021optimizing}. 
It is the equivalence in the transient dynamics rather than the 
steady state that determines whether the transmission features in the early stages
of a memory-dependent disease outbreak can be properly measured through Markovian 
modeling. Moreover, it is only during the transient phase control and mitigation 
strategies can be effective for preventing a large scale outbreak. Discovering 
when and how a non-Markovian process can be approximated by a Markovian process 
during the transient state is thus of paramount importance. To our knowledge, 
such a ``transient-state equivalence'', where the Markovian and non-Markovian 
transmission models produce similar behaviors over the entire transient 
transmission period, has not been established. In fact, the conditions under 
which the transient equivalence may hold are completely unknown at the present.

In this paper, we present results from a comprehensive study of how memory effects
impact the Markovian estimations in terms of the errors that arise from the 
Markovian hypothesis. We consider both the steady-state and transient-state equivalences
between non-Markovian and Markovian models. We first rigorously show that, in the 
steady state, a memory-dependent non-Markovian spreading process is always 
equivalent to certain Markovian (memoryless) ones. We then turn to the transient states 
and find that an approximate equivalence can still be achieved but only if the 
average generation time matches the average removal time 
in the memory-dependent non-Markovian spreading dynamics.
Qualitatively, the equality of the two times gives rise to a memoryless 
correlation between the infection and removal processes, thereby minimizing the 
impact of any memory effects. We establish that the equality gives the condition 
under which Markovian theory accurately describes memory-dependent transmission. 

One fundamental quantity underlying an epidemic process is the basic 
reproduction number $R_0$. Our theoretical analysis indicates that, when the average generation 
and removal times are equal, the transient-state equivalence between 
memory-dependent and memoryless transmissions will minimize the error of Markovian approach in 
estimating $R_0$ and lead to its accurate epidemic forecasting and prevention 
evaluation. Another finding is that the generation-to-removal time ratio plays a 
decisive role in the accuracy of the Markovian approximation. Specifically, 
if the average generation time is smaller (greater) than 
the average removal time, the Markovian approximate will lead to an 
overestimation (underestimation) of $R_0$ and epidemic forecasting
as well as the errors of the prevention 
evaluation, which can also be verified based on readily 
accessible clinical data of 4 types of real diseases. 

Strikingly, the estimation accuracy is largely determined by the time 
ratio and hardly depends on the particular forms of time distributions or the 
specific values of the average generation and removal times. This property is of great practical 
significance, because it is in general challenging to obtain the detailed distributions of the generation 
and removal times in the early stages of the epidemic~\cite{ganyani2020estimating}, 
but their average values can be reliably estimated even during the transient 
phase~\cite{park2021forward,svensson2007note,klinkenberg2011correlation,te2013estimating,champredon2018equivalence}.
Moreover, based on this property, we have developed a semi-empirical mathematical relationship that 
connects the errors in estimating $R_0$ with the generation-to-removal time ratio. 
This relationship holds practical value as it can be utilized to rectify errors in real-world scenarios.
And the rectification of $R_0$ and epidemic forecasting can be accomplished through our web-based application \cite{Validity_MM}.

Overall, our study establishes a general criterion for modeling memory-dependent 
processes within the context of Markovian frameworks. Once the condition for the existence of a 
transient-state equivalence between Markovian and non-Markovian dynamics is fulfilled, 
epidemic forecasting and prevention evaluation can be carried out using the Markovian model, 
again based solely on the data collected from the transient phase.

\section*{Results}

The overall structure of this work is depicted in Fig.~\ref{fig:illustration}. 
The section titled ``Model'' presents the Model building of the age-stratified Susceptible-Infected-Removed (SIR) spreading (Fig.~\ref{fig:illustration}a), 
highlighting the difference between the Markovian (memoryless) and non-Markovian (memory-dependent) dynamics (Fig.~\ref{fig:illustration}b).
In the ``Dynamical equivalence'' section, we demonstrate the equivalence between  Markovian (memoryless) and non-Markovian (memory-dependent) 
dynamics for steady state and transient dynamics, which will further lead to the accurate description of 
memory-dependent dynamics by the Markovian theory (Fig.~\ref{fig:illustration}c). The section ``Markovian approximation of 
memory-dependent spreading dynamics'' analyzes the errors of the Markovian approach 
in estimating $R_0$, epidemic forecasting, and prevention evaluation (Fig.~\ref{fig:illustration}d).

\subsection*{Model}

We articulate an age-stratified SIR spreading dynamics model, in which the entire
population is partitioned into various age groups with intricate age-specific 
contact rates among them. The distribution of population across different age 
groups is represented by an age distribution vector ($\textbf{\textit{p}}$), with the 
age-dependent contact matrix ($\textbf{\textit{A}}$) quantifying the transmission rates 
between different age groups. Both $\textbf{\textit{p}}$ and $\textbf{\textit{A}}$ can be 
constructed from empirical data~\cite{UN_WPP_2021,prem2017projecting}.
For convenience, to distinguish between the actual dynamical process and its 
theoretical treatment, throughout this paper we use the terms ``memory-dependent'' 
and ``memoryless'' to describe actual spreading processes and Monte Carlo 
simulations, while in various theoretical analyses, the corresponding terms 
are ``non-Markovian'' and ``Markovian.'' 

\begin{figure*}%[tbhp]
\centering
\includegraphics[width=0.8\linewidth]{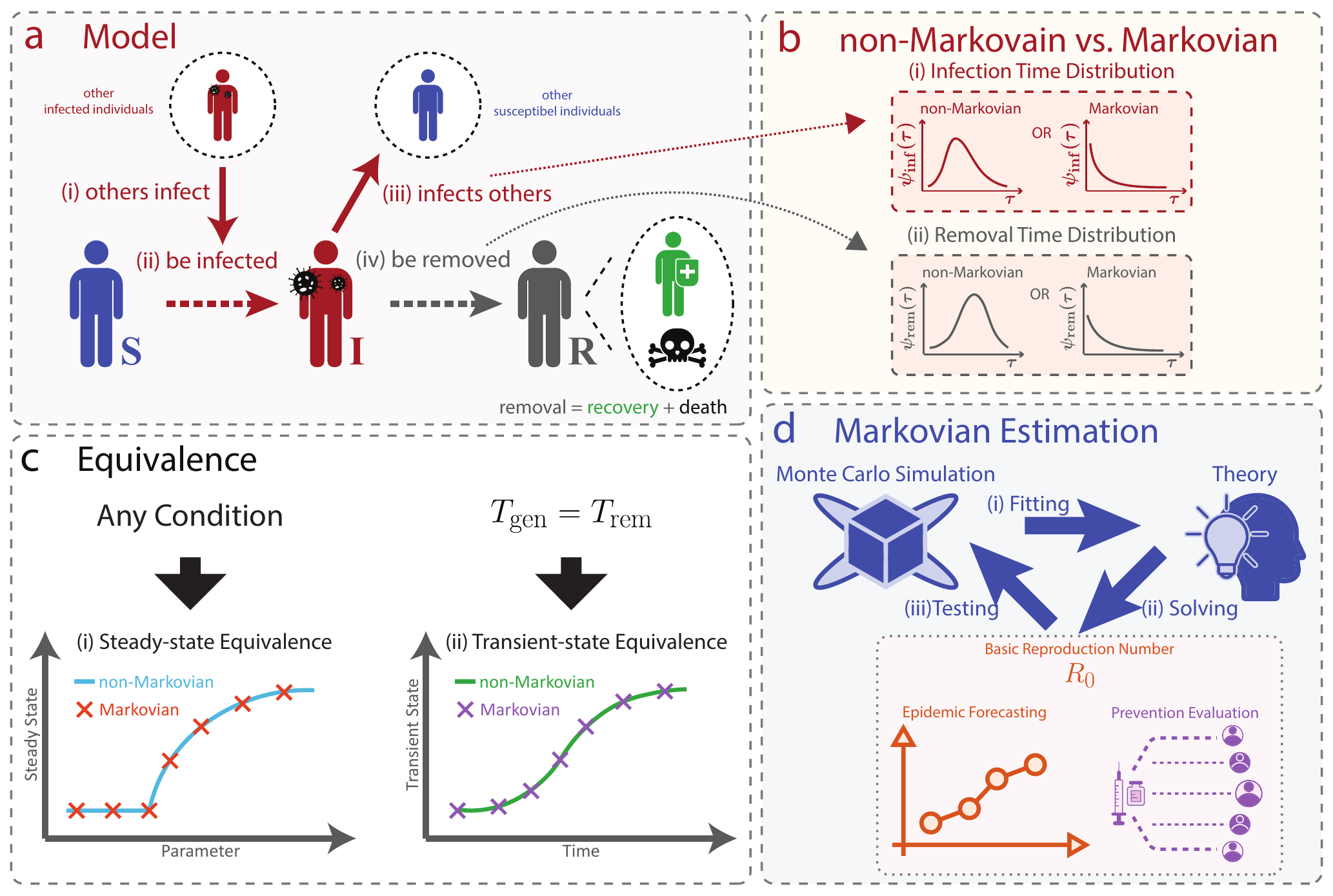}
\caption{\textbf{Overall structure of this work}. 
(\textbf{a}) SIR Model. Each individual belongs to one of the three states: 
susceptible (S), infected (I), or removed (R). When infected (i), a susceptible 
individual will switch into the I state (ii) and gain the ability to infect 
others (iii). An infected individual is removed (through recovery or death) with 
a probability (iv). (\textbf{b}) Non-Markovian versus Markovian process. The 
infection capacity of an infected individual is characterized by the infection 
time distribution $\psi_{\mathrm{inf}}(\tau)$ and its removal can be described by 
the removal time distribution $\psi_{\mathrm{rem}}(\tau)$. For the non-Markovian 
process, the distributions can assume quite general forms, while the distributions
are exponential for a Markovian process. (\textbf{c}) Equivalence between 
non-Markovian and Markovian processes: (i) steady-state equivalence holds under all 
conditions; (ii) transient-state equivalence only holds only when 
$T_{\mathrm{gen}}$ is equal to $T_{\mathrm{rem}}$. (\textbf{d}) Markovian 
estimation of memory-dependent process. (i) The initial phase of the Monte Carlo 
simulation is used to fit the parameters according to 
the Markovian theory. (ii) Important issues such as the estimation of $R_0$, 
epidemic forecasting, and the evaluation of the vaccination strategies can be 
addressed by the theory. (iii) The remaining data generated by the Monte Carlo 
simulation is used to test the accuracy of the estimated $R_0$, epidemic 
forecasting, and prevention evaluation.}
\label{fig:illustration}
\end{figure*}

The mechanism of disease transmission across different age groups and the recovery 
or death of infected individuals can be described by the SIR model, as illustrated 
in Fig.~\ref{fig:illustration}a, where the individuals are placed into three 
compartments: susceptible (S), infected (I), and removed (R). Susceptible 
individuals (S) have not contracted the disease and are at risk of being infected. 
Infected individuals (I) have contracted the disease and can infect others. Removed
individuals (R) have recovered or died from the disease. There are two dynamical 
processes: (1) infection during which susceptible individuals become infected by 
others and transition to the I state so as to become capable of infecting others, 
as shown in Fig.~\ref{fig:illustration}a(i--iii); and (2) removal during which 
infected individuals recover or die from the disease transmission and transition 
to the R state, as shown in Fig.~\ref{fig:illustration}a(iv). The ability to infect
others of an infected individual can be characterized by the infection time 
distribution, $\psi_{\mathrm{inf}}(\tau)$, where $\tau$ denotes the time elapsed 
between the time the individual is infected and the current time, and the 
probability of the infection process occurring during the time interval 
$[\tau,\tau+d\tau)$ is given by $\psi_{\mathrm{inf}}(\tau)d\tau$, as shown in 
Fig.~\ref{fig:illustration}b(i). Likewise, the removal process is described by the 
removal time distribution, $\psi_{\mathrm{rem}}(\tau)$, where the probability of 
a removal occurring within the time interval $[\tau, \tau + d\tau)$ is given by 
$\psi_{\mathrm{rem}}(\tau)d\tau$, as shown in Fig.~\ref{fig:illustration}b(ii). 
The time distributions of the infection and removal processes with memory
effects are general, with the exponential distributions associated with the 
memoryless process being a special case of the memory-dependent process.
    
The generic memory-dependent SIR spreading dynamics can be described by a set of 
deterministic integro-differential equations:
\begin{align} \label{eq:s_t}
	\frac{ds_l(t)}{dt} & = -s_l(t)k\sum_{m=1}^{n}{A_{lm}p_m\int_{0}^{t}{\omega_{\mathrm{inf}}(t-t')\Psi_{\mathrm{rem}}(t-t')dc_m(t')}}, \\ \label{eq:i_t}
	i_l(t) &= \int_{0}^{t}{\Psi_{\mathrm{rem}}(t-t')dc_l(t')}, \\ \label{eq:r_t}
	r_l(t) &= \int_{0}^{t}{[1 - \Psi_{\mathrm{rem}}(t-t')]dc_l(t')},
\end{align} 
where $s_l(t)$, $i_l(t)$, and $r_l(t)$, respectively, denote the fractions of the
susceptible, infected, and removed individuals in age group $l$. The term 
$c_l(t) = 1-s_l(t) = i_l(t)+r_l(t)$ represents the fraction of cumulative 
infections (including both infections and removals) with respect to the total 
population in age group $l$, while $k$ is a parameter to adjust the overall 
contacts and $n$ is the total number of age groups. The quantity 
$\omega_{\mathrm{inf}}(\tau)$ represents the hazard function of 
$\psi_{\mathrm{inf}}(\tau)$, meaning the rate at which infection happens at $\tau$, 
given that the infection has not occurred before $\tau$. $\Psi_{\mathrm{rem}}(\tau)$ is the survival 
function of $\psi_{\mathrm{rem}}(\tau)$, meaning probability that the removal 
has not occurred by $\tau$ (see {\bf Method} for detailed calculations for hazard and survival functions). 
When the infection and removal time 
distributions are known, Eqs.~(\ref{eq:s_t}--\ref{eq:r_t}) provide an accurate 
description of generic (including memory-dependent and memoryless) SIR spreading 
Monte Carlo simulations in an age-stratified population-based system. As shown in 
Figs.~\ref{fig:theories}a--c, the theory is validated by the agreement between 
the numerical solutions of Eqs.~(\ref{eq:s_t}--\ref{eq:r_t}) and the results 
from direct Monte Carlo simulations. (See {\bf Method} for a detailed description
of the Monte Carlo simulation procedure.)

Eqs.~(\ref{eq:s_t}--\ref{eq:r_t}) provide a general framework encompassing 
both non-Markovian and Markovian descriptions. If the infection and removal time 
distributions are exponential: $\psi_{\mathrm{inf}}(\tau) = \gamma e^{-\gamma\tau}$
and $\psi_{\mathrm{rem}}(\tau) = \mu e^{-\mu\tau}$, 
Eqs.~(\ref{eq:s_t}--\ref{eq:r_t}) can be reformulated into a Markovian theory 
and further simplified into a set of ordinary differential equations with the 
constant infection and removal rates $\gamma$ and $\mu$ (A detailed derivation of these equations is presented in Supplementary Note 1) : 
\begin{align} \label{eq:s_t_m}
\frac{ds_l(t)}{dt} &= -s_l(t)k\gamma \sum_{m=1}^{n}{A_{lm}p_mi_m(t)}, \\ \label{eq:i_t_m}
\frac{di_l(t)}{dt} &= s_l(t)k\gamma \sum_{m=1}^{n}{A_{lm}p_mi_m(t)} - \mu i_l(t), \\ \label{eq:r_t_m}
	\frac{dr_l(t)}{dt} &= \mu i_l(t).
\end{align}

While the non-Markovian and Markovian theories [Eqs.~(\ref{eq:s_t}--\ref{eq:r_t})
and Eqs.~(\ref{eq:s_t_m}--\ref{eq:r_t_m}), respectively] describe the 
memory-dependent and memoryless Monte Carlo simulations, we focus on whether the 
Markovian theory can accurately capture memory-dependent dynamics and how memory 
effects influence its accuracy. For this purpose, we seek to establish the 
equivalence between Markovian and non-Markovian approaches for describing 
spreading dynamics.

\subsection*{Dynamical equivalence}
    
Eqs.~(\ref{eq:s_t}--\ref{eq:r_t_m}) provide a base to study the steady-state
equivalence and transient-state equivalence between non-Markovian and Markovian 
theories, where a steady state characterizes the long-term dynamics of disease 
spreading and a transient state is referred to as the short-term behavior prior to 
system's having reached the steady state. As illustrated in Fig.~\ref{fig:illustration}c, note that steady-state equivalence means 
that the two types of spreading dynamics attain identical steady 
states~\cite{starnini2017equivalence,feng2019equivalence,cator2013susceptible},
whereas transient-state equivalence implies that two types of dynamics are
consistent throughout the entire transmission period. Transient-state equivalence
thus implies steady-state equivalence, but not vice versa. Since 
Eqs.~(\ref{eq:s_t}--\ref{eq:r_t_m}) also provides a numerical framework for 
Monte Carlo simulations, the terms ``steady-state equivalence'' and 
``transient-state equivalence'' not only describe the connection between 
non-Markovian and Markovian theories, but also illustrate the relationship between 
memory-dependent and memoryless processes. Therefore, the equivalence between the 
two theories implies the equivalence between the two corresponding processes, 
and vice versa.
    
\subsubsection*{Steady-state equivalence}
        
Eqs.~(\ref{eq:s_t}--\ref{eq:r_t}) give the following transcendental equation
for determining the steady state (see Supplementary None 2 for detailed derivation):
\begin{align} \label{eq:steady_state}
\tilde{s}_l = \acute{s}_le^{-\frac{R_0}{\Lambda_{\mathrm{max}}}\sum_{m=0}^{n}{kA_{lm}p_m(\tilde{r}_m - \acute{r}_m)}},
\end{align} 
where $\tilde{s}_l = \lim\limits_{t\to+\infty}{s_l(t)}$ and 
$\tilde{r}_l = \lim\limits_{t\to+\infty}{r_l(t)}$ denote the fractions of the 
susceptible and removed individuals in age group $l$ at the steady state (note that $\tilde{s}_l = 1 - \tilde{r}_l$, because at steady state, no infection exists), while 
$\acute{s}_l=s_l(0)$ and $\acute{r}_l=r_l(0)$ represent the initial fractions of
the susceptible and removed individuals in this age group. For non-Markovian 
dynamics, basic reproduction number $R_0$ can be determined by:
\begin{align} \label{eq:r0_generic}
R_0 = \Lambda_{\mathrm{max}}\int_{0}^{+\infty}{\omega_{\mathrm{inf}}(\tau)\Psi_{\mathrm{rem}}(\tau)d\tau}.
\end{align}
For Markovian dynamics, $R_0$ is given by
\begin{align} \label{eq:r0_markovian}
R_0 = \frac{\gamma\Lambda_{\mathrm{max}}}{\mu}.
\end{align}
where $\Lambda_{\mathrm{max}}$ is the maximum eigenvalue of the matrix 
$k\textbf{\textit{A}}\circ\textbf{\textit{p}}$, and $\circ$ denotes 
a row-wise Hadamard product between a matrix and a vector (see Supplementary Note 2 for a detailed 
description). Since Eq.~\eqref{eq:steady_state} applies to both non-Markovian and 
Markovian dynamics, an identical $R_0$ value in the two cases will result in 
equivalent steady states from the same initial conditions. Consequently, for a 
given non-Markovian spreading process, there exists an infinite number of 
Markovian models with the same steady state, as the $R_0$ value is only determined by 
the ratio of $\gamma$ to $\mu$, but not by either value.

\begin{figure*}%[tbhp]
\centering
\includegraphics[width=0.9\linewidth]{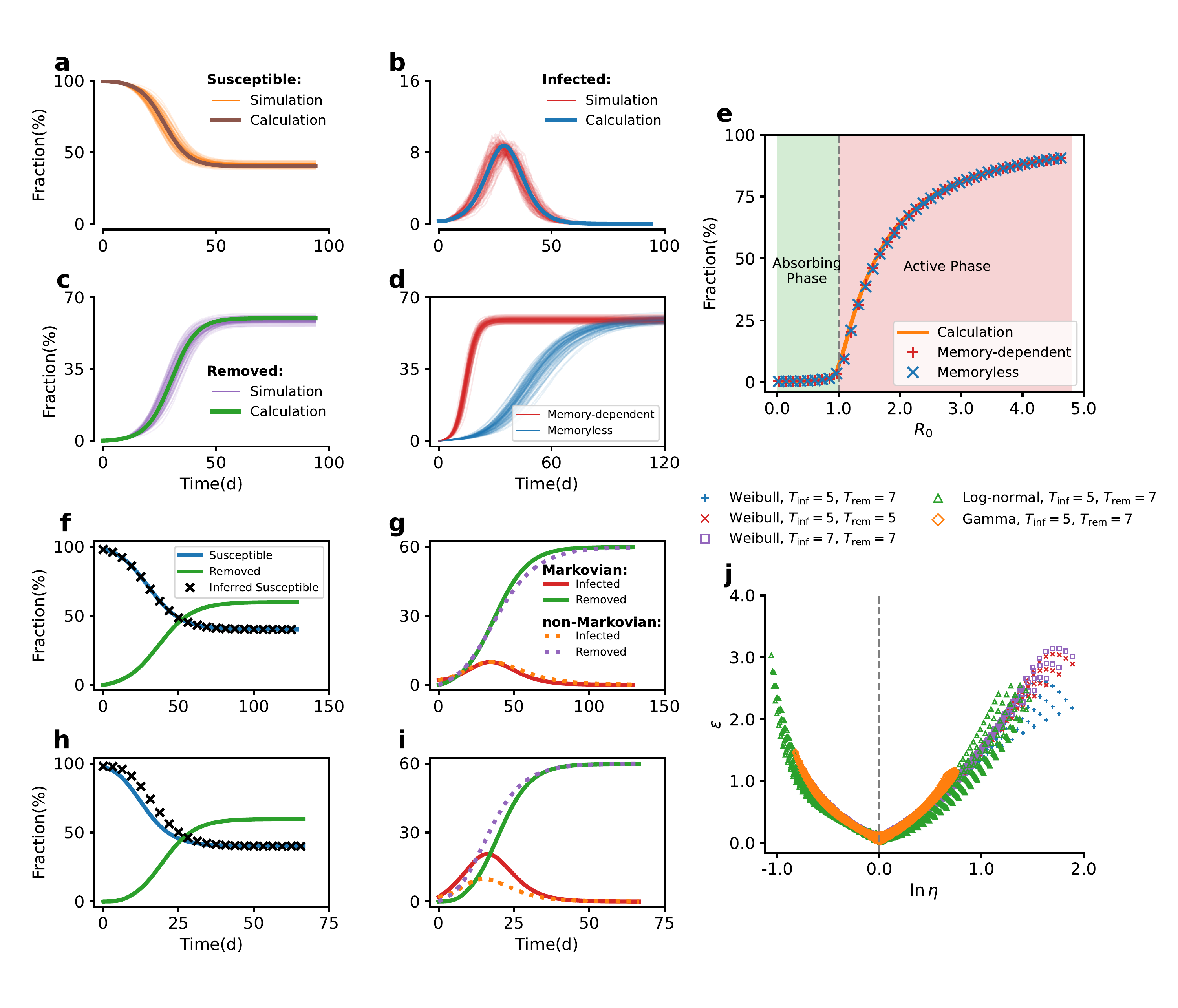}
\caption{\textbf{Steady-state and transient-state equivalence between Markovian
and non-Markovian dynamics}. \textbf{a--c}: The solid brown, blue, and green 
curves represent the theoretical results of the susceptible, infected, and removed
fractions, while the solid orange, red, and purple curves show the corresponding 
results of 100 independent Monte Carlo simulations. \textbf{d}: The red and blue 
curves, respectively, depict the removed fractions from the memory-dependent and 
memoryless Monte Carlo simulations of 100 independent realizations with 
steady-state equivalence. \textbf{e}: Red $+$ and blue $\times$ markers, 
respectively, represent the steady-state removed fractions of memory-dependent 
and memoryless Monte Carlo simulations for different values of $R_0$, where each 
marker is the result of averaging 100 independent simulations. The orange curve 
is the numerical calculations from Eq.~\eqref{eq:steady_state}, and the vertical 
dashed line denotes the critical point $R_0=1$. \textbf{f--g}: For 
$T_{\mathrm{gen}} = T_{\mathrm{rem}}$ in the non-Markovian theory: the blue and 
green curves in \textbf{f} denote the susceptible and removed fractions, while 
the black $\times$ markers represent the inferred susceptible fractions calculated
by substituting removed fractions in Eq.~\eqref{eq:transient_approximation}, which
agrees with the susceptible curve calculated from Eq.~\eqref{eq:s_t}. The red and blue curves in \textbf{g} denote 
the non-Markovian susceptible and removed fractions, while the orange and purple 
dashed curves are the corresponding curves of the Markovian transmission obtained 
from Eqs.~(\ref{eq:inferred_gamma}--\ref{eq:inferred_mu}), which agree with the
non-Markovian results. (The Euler-Lotka equation 
assumes exponential growth of a disease outbreak during the initial stage. 
As a result, the Markovian curves in \textbf{g} slightly deviate from the non-Markovian 
ones as the cumulative infections increase.) \textbf{h--i}: For $T_{\mathrm{gen}} \ne T_{\mathrm{rem}}$ in the non-Markovian theory, the inferred susceptible curve does not match the 
numerical result, and the susceptible and infected curves of the Markovian 
transmission obtained from Eqs.~(\ref{eq:inferred_gamma}--\ref{eq:inferred_mu}) 
do not match the corresponding non-Markovian results. \textbf{j}: Five scenarios 
for the non-Markovian time-distribution setting (within each scenario, $T_\mathrm{inf}$ and $T_\mathrm{rem}$ are fixed): Weibull, $T_{\mathrm{inf}}=5$, 
$T_{\mathrm{rem}}=7$ (blue $+$); Weibull, $T_{\mathrm{inf}}=5$, 
$T_{\mathrm{rem}}=5$ (red $\times$); Weibull, $T_{\mathrm{inf}}=7$, 
$T_{\mathrm{rem}}=7$ (purple $\Box$); log-normal, $T_{\mathrm{inf}}=5$, 
$T_{\mathrm{rem}}=7$ (green $\triangle$); gamma, $T_{\mathrm{inf}}=5$, 
$T_{\mathrm{rem}}=7$ (orange $\Diamond$). The value of $T_{\mathrm{gen}}$ is
modified to adjust $\ln{\eta}$ for better visualization.}
\label{fig:theories} 
\end{figure*}

As shown in Fig.~\ref{fig:theories}d, memory-dependent and memoryless spreading
dynamics that reach the same steady state with the identical $R_0$ value confirm 
the steady-state equivalence. Fig.~\ref{fig:theories}e demonstrates that, even 
for $R_0$ ranging from $0.01$ to $2$, the equivalent memory-dependent and 
memoryless spreading dynamics still produce highly consistent steady states that 
can be calculated from Eq.~\eqref{eq:steady_state}, which share the same critical 
point of phase transition at $R_0 = 1$.
        
\subsubsection*{Transient-state equivalence}

From the preceding section, we used the basic reproduction number $R_0$, a fundamental 
metric quantifying the number of secondary infections generated by a single 
individual, to characterize the steady-state equivalence. Here, we propose to 
quantify the transient-state equivalence through the average generation time
$T_{\mathrm{gen}}$ that measures the ``velocity'' at which secondary infections 
occur. This time can be calculated 
as~\cite{ferretti2020quantifying,wallinga2007generation}, 
\begin{align} \nonumber
T_{\mathrm{gen}} = \int_{0}^{+\infty}{\tau\psi_{\mathrm{gen}}(\tau)d\tau}, 
\end{align}
where 
\begin{align}
\psi_{\mathrm{gen}}(\tau) = \frac{\omega_{\mathrm{inf}}(\tau)\Psi_{\mathrm{rem}}(\tau)}{\int_{0}^{+\infty}{\omega_{\mathrm{inf}}(\tau')\Psi_{\mathrm{rem}}(\tau')d\tau'}}
\label{eq:psi_gen}
\end{align}
is the generation time distribution. Effectively, $T_{\mathrm{gen}}$ measures the 
average duration of disease transmission from an infected individual to the next 
generation of individuals. Likewise, the average infection time $T_{\mathrm{inf}}$ 
and the average removal time $T_{\mathrm{rem}}$ are defined as the mean values of 
the infection and removal time distributions: 
\begin{align} \nonumber
T_{\mathrm{inf}} & = \int_{0}^{+\infty}{\tau\psi_{\mathrm{inf}}(\tau)d\tau}, \\ \nonumber
T_{\mathrm{rem}} & = \int_{0}^{+\infty}{\tau\psi_{\mathrm{rem}}(\tau)d\tau}.
\end{align}
In calculating $T_{\mathrm{gen}}$, the individual's removal is taken into 
account, while $T_{\mathrm{inf}}$ measures the average time of the first disease 
transmission of an infectious individual without factoring in removal. In the 
classical memoryless transmission with exponential distributions 
$\psi_{\mathrm{inf}}(\tau)$ and $\psi_{\mathrm{rem}}(\tau)$, the equality 
$T_{\mathrm{gen}} = T_{\mathrm{rem}}$ holds. However, for memory-dependent 
spreading, the possible scenarios are: $T_{\mathrm{gen}} = T_{\mathrm{rem}}$, 
$T_{\mathrm{gen}} < T_{\mathrm{rem}}$, or $T_{\mathrm{gen}} > T_{\mathrm{rem}}$.
Specifically, because $T_{\mathrm{gen}}$, $T_{\mathrm{inf}}$ and $T_{\mathrm{rem}}$
all represent the mean values of distributions, it is possible for 
$T_{\mathrm{rem}}$ to be shorter than $T_{\mathrm{gen}}$ or $T_{\mathrm{inf}}$ in 
some situations. And our web-based application demonstrates the impact of parameters 
on the time distributions (infection, removal, and generation) as well as their average times \cite{Validity_MM}.
        
For a non-Markovian spreading process, if the equality 
$T_{\mathrm{gen}} = T_{\mathrm{rem}}$ holds, in the transient state we have
\begin{align} \label{eq:transient_approximation}
s_l(t) \simeq \acute{s}_le^{-\frac{R_0}{\Lambda_{\mathrm{max}}}\sum_{m=0}^{n}{kA_{lm}p_m[r_m(t) - \acute{r}_m]}},
\end{align}
which exhibits a memoryless transmission pattern similar to Markovian dynamics 
(see Supplementary Notes 3 for a detailed analysis). 
Intuitively, the equality $T_{\mathrm{gen}} = T_{\mathrm{rem}}$ signifies that 
the infection and removal processes occur concurrently, which in turn leads to 
a memoryless relationship between the two processes, thereby minimizing the memory 
effects. Furthermore, to determine the 
corresponding Markovian parameters $\gamma$ and $\mu$ of
the Markovian transmission which is
equivalent to the non-Markovian dynamics in the transient state, 
we need to utilize the Euler-Lotka equation~\cite{dietz1993estimation, wallinga2007generation, park2021forward}:
\begin{align}
1 = R_0\int_{0}^{+\infty}{e^{-g\tau}\psi_{\mathrm{gen}}(\tau)d\tau},
\label{eq:euler_lotka}
\end{align}
where $g$ denotes the growth rate of the non-Markovian dynamics and 
is another measure of how quickly the epidemic is spreading within a population.
Therefore, we can calculate the values of the basic reproduction number, $R_0$, 
and growth rate, $g$, in the non-Markovian dynamic
by using Eqs.~\eqref{eq:r0_generic}, \eqref{eq:psi_gen} and \eqref{eq:euler_lotka}. Additionally,
the Markovian form of $\psi_{\mathrm{gen}}(\tau)$ according to Eq.~\eqref{eq:psi_gen} is $\mu e^{-\mu\tau}$, 
and the equivalent Markovian and non-Markovian dynamics 
in the transient state have the same values of $R_0$ and the equal values of $g$.
By substituting $\psi_{\mathrm{gen}}(\tau) = \mu e^{-\mu\tau}$ 
and the calculated $R_0$ and $g$ into Eq.~\eqref{eq:euler_lotka}, 
we can determine the value of $\mu$. Furthermore, using Eq.~\eqref{eq:r0_markovian}, 
we can find the value of $\gamma$ based on $\mu$. Hence, 
the Markovian parameters $\gamma$ and $\mu$ are determined as follows:
\begin{align} \label{eq:inferred_gamma}
\gamma & = \frac{gR_0}{\Lambda_{\mathrm{max}}(R_0 - 1)}, \\ \label{eq:inferred_mu}
\mu & = \frac{g}{R_0 - 1}.
\end{align}
And we provide visualizations that illustrate how the values of $\gamma$ and $\mu$ 
are influenced by the distribution parameters in our web-based application\cite{Validity_MM}.

As illustrated in Figs.~\ref{fig:theories}f--g, when the equality 
$T_{\mathrm{gen}} = T_{\mathrm{rem}}$ holds for the non-Markovian dynamics,
Eq.~\eqref{eq:transient_approximation} holds, which can be seen by comparing the 
susceptible curve calculated from Eq.~\eqref{eq:s_t} to that inferred from 
Eq.~\eqref{eq:transient_approximation}, as shown in Fig.~\ref{fig:theories}f. 
In this case, the Markovian spreading curves deduced from 
Eqs.~(\ref{eq:inferred_gamma}--\ref{eq:inferred_mu}) closely align with the 
non-Markovian transient curves, as shown in Fig.~\ref{fig:theories}g. However, 
as shown in Figs.~\ref{fig:theories}h--i, if the equality does not hold, the 
equivalence in transient states breaks down. It is important to note that the 
Euler-Lotka equation assumes an exponential growth of a disease outbreak and 
is only reasonable 
at the initial stage. Consequently, as the cumulative infections increase (Fig.~\ref{fig:theories}g), 
the Markovian curves will exhibit slight deviations from the non-Markovian counterparts.
Meanwhile, because the equivalent dynamics share the same $R_0$, they will ultimately reach the same steady state, ensuring that deviations will diminish while they approach the steady state.

To evaluate, under different values of the generation-to-removal time ratio 
$\eta \equiv T_{\mathrm{gen}}/T_{\mathrm{rem}}$ for non-Markovian dynamics 
we introduce a metric, $\varepsilon$, to quantify
the difference from the corresponding Markovian results calculated from
Eqs.~(\ref{eq:inferred_gamma}--\ref{eq:inferred_mu}) (see {\bf Method} for detailed definition of $\varepsilon$).
Fig.~\ref{fig:theories}j shows, for non-Markovian numerical calculations, five 
scenarios under various forms of time distributions $\psi_{\mathrm{inf}}(\tau)$
and $\psi_{\mathrm{rem}}(\tau)$ constrained by certain average infection and 
removal times: Weibull, $T_{\mathrm{inf}} = 5$, $T_{\mathrm{rem}} = 7$; Weibull, 
$T_{\mathrm{inf}} = 7$, $T_{\mathrm{rem}} = 7$; Weibull, $T_{\mathrm{inf}} = 5$, 
$T_{\mathrm{rem}} = 5$; log-normal, $T_{\mathrm{inf}} = 5$, $T_{\mathrm{rem}} = 7$;
and gamma, $T_{\mathrm{inf}} = 5$, $T_{\mathrm{rem}} = 7$ 
(see {\bf Method} for detailed definitions of Weibull, log-normal, and gamma distributions). The $T_{\mathrm{gen}}$ 
value is adjusted to obtain different values of $\ln{\eta}$. For $\ln{\eta}=0$, 
i.e., $T_{\mathrm{gen}} = T_{\mathrm{rem}}$, the ``distance'' $\varepsilon$ 
between the transient states of non-Markovian and Markovian dynamics with 
parameters determined from Eqs.~(\ref{eq:inferred_gamma}--\ref{eq:inferred_mu}) 
is minimal. Otherwise, $\varepsilon$ increases as $\ln{\eta}$ deviates from zero. 
Remarkably, Fig.~\ref{fig:theories}j shows that $\varepsilon$ depends only on the 
ratio of $T_{\mathrm{gen}}$ to $T_{\mathrm{rem}}$, not on the values of 
$T_{\mathrm{gen}}$, $T_{\mathrm{inf}}$, or $T_{\mathrm{rem}}$, and it is not 
affected by the specific form of the time distributions.

Furthermore, it is important to note that the condition 
$T_{\mathrm{gen}} = T_{\mathrm{rem}}$ can guarantee transient-state equivalence 
between a non-Markovian dynamic and a Markovian one, 
but according to Eqs.~(\ref{eq:inferred_gamma}--\ref{eq:inferred_mu}), 
it does not imply that the average generation and removal times of the 
non-Markovian dynamic must be equal to those of the equivalent Markovian one. 
For instance, if a non-Markovian dynamic satisfies the condition of transient-state 
equivalence and we keep its average generation and removal times fixed, 
altering the shape of the corresponding time distributions will change 
the transmission speed \cite{wallinga2007generation}. This change, in turn, affects the infection and 
removal rates of the equivalent Markovian dynamic, leading to different 
average generation and removal times for the Markovian equivalent dynamic
(see Supplementary 4 and 5 for a detailed analysis).

\subsection*{Markovian approximation of memory-dependent spreading dynamics}
   
As illustrated in Fig.~\ref{fig:illustration}d, testing the applicability of 
Markovian theory for memory-dependent spreading dynamics requires three steps.
The first step is fitting, where the memory-dependent Monte Carlo simulation data 
are divided into two parts: (a) a short initial period used as the training data 
for fitting the Markovian parameters in Eqs.~(\ref{eq:s_t_m}--\ref{eq:r_t_m}), 
and (b) the remaining testing data for evaluating the performance of the 
Markovian model (see {\bf Methods} for details of the fitting procedure). 
The second step is to use the fitted Markovian model for tasks such as estimating 
$R_0$, predicting outbreaks and assessing the prevention effects of different 
vaccination strategies. The third step is testing, i.e., evaluating the accuracy 
of the Markovian model, e.g., by comparing the estimated and actual 
$R_0$ values, disease outbreaks and prevention effects. As real-world disease spreading is subject to environmental, social and 
political disturbing factors, for the fitting and testing steps, we conduct Monte 
Carlo simulations of stochastic memory-dependent disease outbreaks to generate the 
training and testing data. 

Here, we first analyze the influence of $\eta$ on the estimation of $R_0$ using
the Markovian theory, and use the results to design two tasks to evaluate the 
applicability of the theory in epidemic forecasting and prevention evaluation of 
memory-dependent spreading. For comparison, we also generate the corresponding 
results from the non-Markovian theory in the two tasks.
    
\subsubsection*{Estimation of basic reproduction number}

\begin{figure*}%[tbhp]
\centering
\begin{subfigure}
\centering
\includegraphics[width=0.9\linewidth]{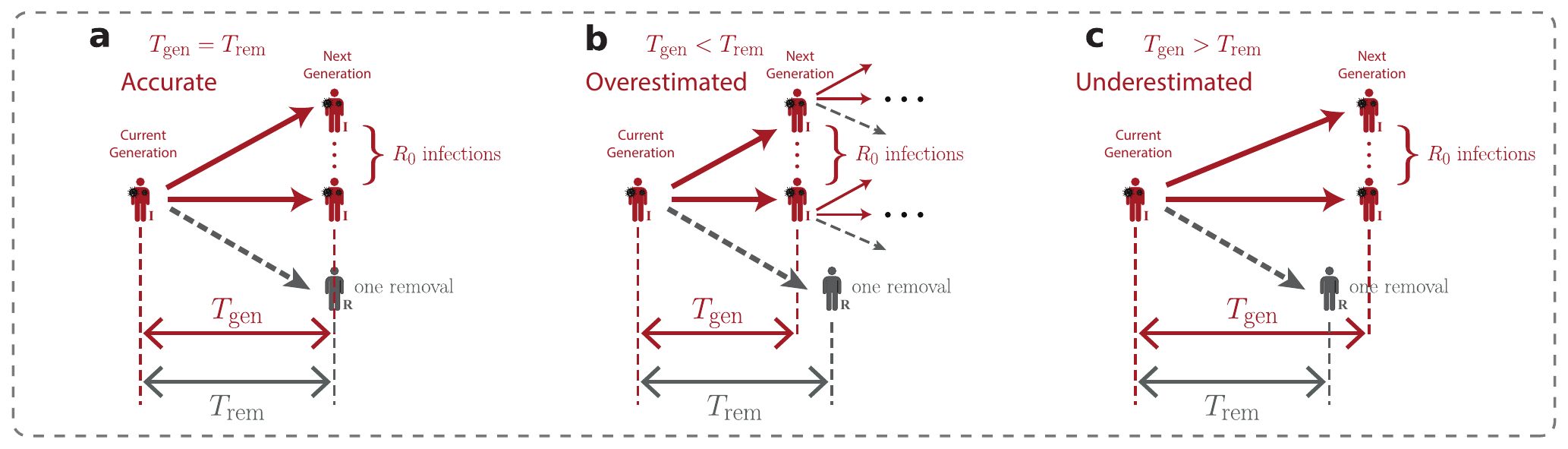}
\end{subfigure}
\hfill
\begin{subfigure}
\centering
\includegraphics[width=0.9\linewidth]{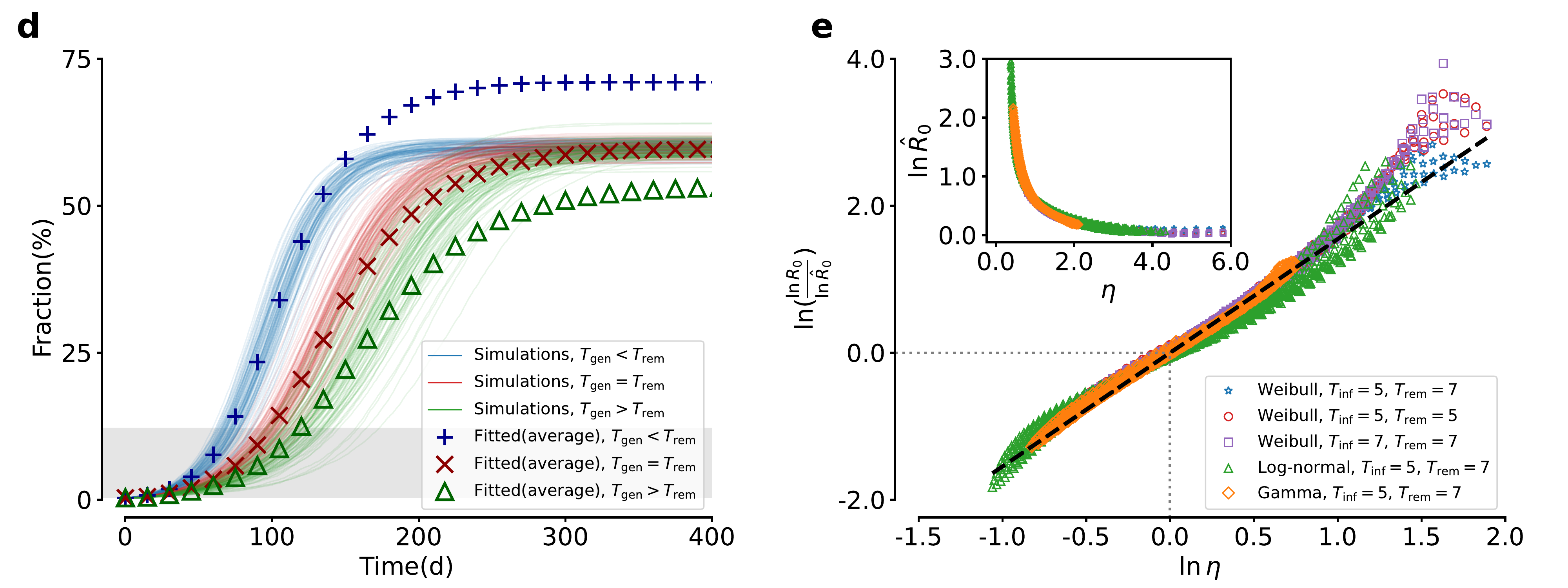}
\end{subfigure}
\caption{\textbf{Estimation of $R_0$}. 
\textbf{a}--\textbf{c}: Mechanism of the $R_0$ estimation. 
The red arrows represent the infection process of the next generation by the 
current generation, while the dashed arrows denote the removal of the current 
generation. The relationship between $T_{\mathrm{gen}}$ and $T_{\mathrm{rem}}$ 
influences the number of new infections when the current generation of infections 
is removed. For $T_{\mathrm{gen}} = T_{\mathrm{rem}}$, the number of new 
infections is exactly $R_0$. For $T_{\mathrm{gen}} < T_{\mathrm{rem}}$, the 
number of new infections is greater than $R_0$. For 
$T_{\mathrm{gen}} > T_{\mathrm{rem}}$, the number of new infections is smaller 
than $R_0$. \textbf{d}: Three distinct categories of
disease spreading with the same value of $R_0$: 
$T_{\mathrm{gen}} < T_{\mathrm{rem}}$ (blue curves), 
$T_{\mathrm{gen}} = T_{\mathrm{rem}}$ (red curves), and 
$T_{\mathrm{gen}} > T_{\mathrm{rem}}$) (green curves), where the fractions of 
cumulative infected individuals (i.e., sum of infected and removed fractions) are calculated using 100 independent realizations. The 
predicted future evolution of the spreading dynamics by the Markovian theory with 
the fitted parameters are also shown:  
$T_{\mathrm{gen}} < T_{\mathrm{rem}}$ (blue $+$ symbols),
$T_{\mathrm{gen}} = T_{\mathrm{rem}}$ (red $\times$ symbols),
and $T_{\mathrm{gen}} > T_{\mathrm{rem}}$ (green $\triangle$ symbols),
where the gray area marks the average cumulative infected fraction for selecting 
the training data (see {\bf Method} for details).
\textbf{e}: The relationship between $\ln{(\ln{R_0}/\ln{\hat{R}_0})}$ ($R_0$ represents the 
real basic reproduction number, while $\hat{R_0}$ denoted the estimated one) and $\ln{\eta}$ 
where the horizontal and vertical dotted lines show that the equality between
$T_{\mathrm{gen}}$ and $T_{\mathrm{rem}}$ results in an accurate estimation of 
$R_0$ and the dashed line represents a linear fitting with the slope $1.55$. 
Inset: the relation between $\ln\hat{R}_0$ and $\eta$ with the asymptotic behaviors: 
for $\eta\to 0$, $\hat{R}_0 \to +\infty$ ($\ln{\hat{R}_0} \to +\infty$), and for $\eta\to +\infty$,
$\hat{R}_0 \to 1$ ($\ln{\hat{R}_0} \to 1$).}
\label{fig:analysis} 
\end{figure*}

Estimating basic reproduction number $R_0$ is crucial for determining the ultimate prevalence of disease 
spreading and for assessing the effectiveness of various disease containment 
measures~\cite{dietz1993estimation,wallinga2007generation,park2021forward}. When
using the Markovian theory to fit the early-stage transmission of a 
memory-dependent process, a key parameter that can affect the estimation of 
$R_0$ is the ratio $\eta$. To develop an analysis, recall the basic principle
for estimating $R_0$: disease spreading dynamics can be viewed as a combination 
of two parallel processes: infection and removal. In particular, the infection 
process is the reproduction of the disease within each generation, where each 
infected individual generates an average of $R_0$ newly infected individuals in 
the subsequent generation after a mean time period $T_{\mathrm{gen}}$. In the
removal process, infected individuals are removed from the spreading chain, where 
each generation takes an average time $T_{\mathrm{rem}}$ to be removed. For 
a Markovian type of dynamics with constant $\gamma$ and $\mu$, the equality
$T_{\mathrm{gen}} = T_{\mathrm{rem}}$ holds. Consequently, during the Markovian 
fitting step, the average number of new infections upon the removal of a single 
infected individual is taken as the value of $R_0$. For memory-dependent spreading,
if the equality $T_{\mathrm{gen}} = T_{\mathrm{rem}}$ holds, the memory-dependent 
spreading curves will possess an approximate memoryless feature so that $R_0$ 
can be still be estimated by counting the number of new infections at the time 
when the current generation of infections is removed, as shown in  
Fig.~\ref{fig:analysis}a. However, for $T_{\mathrm{gen}}<T_{\mathrm{rem}}$, 
more than one generation is produced while the current generation is removed, 
$R_0$ estimated by the Markovian theory will represent an overestimate, as shown
in Fig.~\ref{fig:analysis}b. For $T_{\mathrm{gen}} > T_{\mathrm{rem}}$, 
less than one generation is created during $T_{\mathrm{rem}}$, the Markovian theory
will give an underestimate of $R_0$, as shown in Fig.~\ref{fig:analysis}c.

Fig.~\ref{fig:analysis}d shows fitting curves (training data) from the early 
stage of memory-dependent spreading simulations that have identical $R_0$ values 
for $T_{\mathrm{gen}} = T_{\mathrm{rem}}$ (red curves), 
$T_{\mathrm{gen}} < T_{\mathrm{rem}}$ (blue curves), 
and $T_{\mathrm{gen}} > T_{\mathrm{rem}}$ (green curves) from the Markovian 
theory. When the equality $T_{\mathrm{gen}} = T_{\mathrm{rem}}$ holds, the 
Markovian theory with fitted parameters generates accurate the future evolution 
(red $\times$ symbols). For $T_{\mathrm{gen}} < T_{\mathrm{rem}}$, the outbreak 
in the initial stage is accelerated, resulting in an overestimation by the 
Markovian theory (blue $+$ symbols). For $T_{\mathrm{gen}} < T_{\mathrm{rem}}$,
the initial outbreaks are decelerated, leading to an underestimation by the 
Markovian approach (blue $+$ symbols).

The above qualitative insights lead to a semi-empirical relationship between the 
Markovian-estimated basic reproduction number $\hat{R}_0$ and its actual value 
$R_0$ as:
\begin{align} \label{eq:r0_estimation}
\hat{R}_0 = (R_0)^{\eta^{-a}},
\end{align}
where $a$ is a positive coefficient (see {\bf Method} for a detailed 
derivation). The value of $a$ is a crucial and constant parameter in Eq.~\eqref{eq:r0_estimation}, 
and it needs to be determined by fitting it to the data. Once this constant $a$ is obtained, 
the actual value of $R_0$ can be derived by adjusting the estimated $\hat{R}_0$ based on Eq.~\eqref{eq:r0_estimation},
and more accurate steady state can be calculated by using Eq.~\eqref{eq:steady_state}.

Eq.~\eqref{eq:r0_estimation} implies the relationship 
$\ln{(\ln{R_0}/\ln{\hat{R}_0})} = a\ln{\eta}$. We use the five scenarios specified 
in Fig.~\ref{fig:theories}j for memory-dependent Monte Carlo simulations. 
Fig.~\ref{fig:analysis}e shows the linear relationship between
$\ln{(\ln{R_0}/\ln{\hat{R}_0})}$ and $\ln{\eta}$, providing support for 
our qualitative analysis of the Markovian estimation. The estimation of $R_0$ 
also depends on the ratio $\eta$ and is relatively insensitive to the particular 
forms of the time distributions or the specific values of $T_{\mathrm{gen}}$, 
$T_{\mathrm{inf}}$, or $T_{\mathrm{rem}}$. The results in the inset of 
Fig.~\ref{fig:analysis}e further confirm that the estimated $\hat{R}_0$ approaches 
one when $T_{\mathrm{gen}}$ is much larger than $T_{\mathrm{rem}}$ and tends to 
$+\infty$ when $T_{\mathrm{gen}}$ is much smaller than $T_{\mathrm{rem}}$.
By fitting the available data, we have determined the value of $a$ to be $1.55$.
After obtaining the value of $a$, we can develop our web-based application for rectifying
$R_0$ and epidemic forecasting~\cite{Validity_MM}.

\subsubsection*{Epidemic forecasting}

As suggested in Fig.~\ref{fig:illustration}d, we evaluate the efficacy of Markovian
theory for epidemic forecasting. We use the initial period of Monte Carlo 
simulation data to fit parameters under both Markovian and non-Markovian 
hypotheses and then to predict future disease outbreaks. The remaining simulation 
data are leveraged to evaluate the accuracy of the Markovian and non-Markovian 
forecasting results. Regardless of the type of time distributions in the 
memory-dependent Monte Carlo simulations (Weibull, log-normal, or gamma),
the non-Markovian model fits the training data in a consistent manner, 
i.e., by selecting Weibull time distribution.

%To forecast the dynamical evolution or to evaluate the prevention strategy, 
%we use the Weibull time distributions as the prior and posterior distributions. 

\begin{figure*}%[tbhp]
\centering
\includegraphics[width=\linewidth]{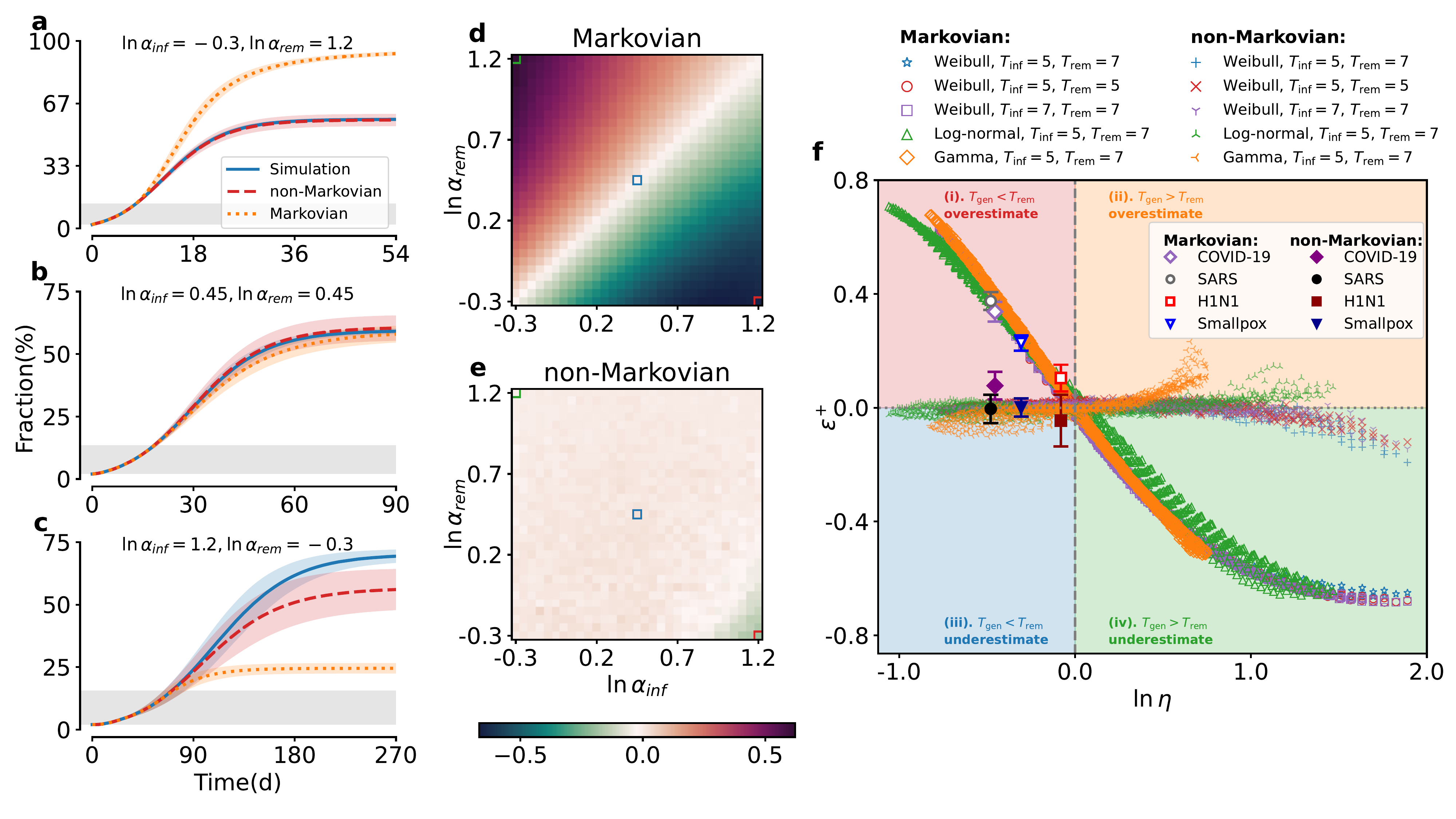}
\caption{\textbf{Epidemic Forecasting}. \textbf{a--c}: Predicted evolution of
the cumulative infected fraction (i.e., sum of infected and removed fractions) by the Markovian (orange dotted curves) and non-Markovian 
(red dashed curves) theories, in comparison with the Monte Carlo simulations with 
Weibull time distributions (blue solid curves), for three sets of simulation 
parameters, respectively. The results are the averages of 100 independent 
realizations with the standard deviations indicated by the shaded regions.
The gray area marks the average cumulative infected fraction for training data 
selection. \textbf{d-e}: The forecasting errors $\varepsilon^{+}$ of Markovian 
and non-Markovian theories with respect to the memory-dependent Monte Carlo 
simulations in the parameter plane of $\ln{\alpha_{\mathrm{inf}}}$ and $\ln{\alpha_{\mathrm{rem}}}$ 
in the range $[-0.3,1.2]$. The green, blue and red squares mark the parameters of 
Monte Carlo simulations in \textbf{a}--\textbf{c}, respectively. \textbf{f}: The 
forecasting errors $\varepsilon^{+}$ from the Markovian and non-Markovian 
theories for different values of $\ln{\eta}$ under five scenarios of time 
distribution setting for Monte Carlo simulations. The corresponding estimations 
for a number of real-world diseases (COVID-19, SARS, H1N1 and Smallpox) are also 
included.}
\label{fig:forecasting}
\end{figure*}

Figs.~\ref{fig:forecasting}a--c show the evolution of the spreading dynamics
from three types of memory-dependent Monte Carlo simulations with Weibull infection
and removal distributions, where the shape parameters $\alpha_{\mathrm{inf}}$ and 
$\alpha_{\mathrm{rem}}$ are selected according to 
$\ln{\alpha_{\mathrm{inf}}} = -0.3, \ln{\alpha_{\mathrm{rem}}} = 1.2$ 
(Fig.~\ref{fig:forecasting}a), $\ln{\alpha_{\mathrm{inf}}} = 0.45, 
\ln{\alpha_{\mathrm{rem}}} = 0.45$ (Fig.~\ref{fig:forecasting}b), and 
$\ln{\alpha_{\mathrm{inf}}} = 1.2, \ln{\alpha_{\mathrm{rem}}} = -0.3$ 
(Fig.~\ref{fig:forecasting}c), for $T_{\mathrm{inf}}=5$ and $T_{\mathrm{rem}}=7$.
For the Weibull distributions, we have 
$\alpha_{\mathrm{inf}}<\alpha_{\mathrm{rem}}$, 
$\alpha_{\mathrm{inf}} = \alpha_{\mathrm{rem}}$ and 
$\alpha_{\mathrm{inf}} > \alpha_{\mathrm{rem}}$, corresponding to 
$T_{\mathrm{gen}} < T_{\mathrm{rem}}$, $T_{\mathrm{gen}} = T_{\mathrm{rem}}$, 
and $T_{\mathrm{gen}} > T_{\mathrm{rem}}$, respectively. We compare the simulated 
cumulative infected fractions to those predicted by the Markovian and non-Markovian
theories. In general, the non-Markovian theory provides more accurate predictions 
than the Markovian theory. For the specific parameter setting 
$\ln{\alpha_{\mathrm{inf}}} = 0.45, \ln{\alpha_{\mathrm{rem}}} = 0.45$ 
(i.e., $T_{\mathrm{gen}} = T_{\mathrm{rem}}$), both theories yield a high accuracy.

The accuracy can be assessed through the forecasting error $\varepsilon^{+}$ that 
evaluates whether a theory overestimates or underestimates the steady-state 
cumulative infection, i.e., quantifying the extent of deviation between the results 
obtained from Markovian or non-Markovian theories and those derived from Monte Carlo simulations
(see {\bf Method} for detailed definition of $\varepsilon^{+}$). A plus value of $\varepsilon^{+}$ means overestimation 
while minus value indicates underestimation. We evaluate the accuracy measure 
$\varepsilon^{+}$ in the parameter plane of $\ln{\alpha_{\mathrm{inf}}}$ and 
$\ln{\alpha_{\mathrm{rem}}}$, ranging from $-0.3$ to $1.2$. 
Figs.~\ref{fig:forecasting}d--e show that the Markovian accuracy is sensitive 
to parameter changes: underestimated if $\alpha_{\mathrm{inf}}$ is greater than 
$\alpha_{\mathrm{rem}}$ ($T_{\mathrm{gen}} > T_{\mathrm{rem}}$), overestimated 
when $\alpha_{\mathrm{inf}}$ is smaller than $\alpha_{\mathrm{rem}}$ 
($T_{\mathrm{gen}} < T_{\mathrm{rem}}$), and a high forecasting accuracy is 
achieved only for $\alpha_{\mathrm{inf}} = \alpha_{\mathrm{rem}}$ 
($T_{\mathrm{gen}} = T_{\mathrm{rem}}$). In contrast, the non-Markovian theory 
yields highly accurate results in the whole parameter plane, with only a slight 
underestimation for $\alpha_{\mathrm{inf}} \gg \alpha_{\mathrm{rem}}$ 
($T_{\mathrm{gen}} \gg T_{\mathrm{rem}}$, this is primarily due to the increased 
difficulty in fitting simulation data, as the simulation parameters become increasingly unreasonable.).

Using the five scenarios specified in Fig.~\ref{fig:theories}j for memory-dependent 
Monte Carlo simulations, we obtain the relationship between $\varepsilon^{+}$ and 
$\ln{\eta}$, as shown in Fig.~\ref{fig:forecasting}f. It can be seen that, in the 
Markovian framework, an overestimation arises for 
$T_{\mathrm{gen}} < T_{\mathrm{rem}}$, and an underestimation occurs for 
$T_{\mathrm{gen}} > T_{\mathrm{rem}}$. Only when 
$T_{\mathrm{gen}} = T_{\mathrm{rem}}$ is an accurate estimate achieved. 
In general, the non-Markovian theory provides much more accurate forecasting 
than the Markovian theory, especially when $T_{\mathrm{gen}}$ and 
$T_{\mathrm{rem}}$ are not equal. The results further illustrate that the forms 
of time distributions or the specific values of $T_{\mathrm{gen}}$, 
$T_{\mathrm{inf}}$, or $T_{\mathrm{rem}}$ have little impact on forecasting
accuracy.
       
To establish the relevance of these results to real-world diseases, we obtain
the $\psi_{\mathrm{inf}}(\tau)$ and $\psi_{\mathrm{rem}}(\tau)$ relations for 
four known infectious diseases, including COVID-19, SARS, H1N1 influenza, and 
smallpox, using the information in 
Refs.~\cite{ferretti2020quantifying,moghadas2020implications,lauer2020incubation,gatto2020spread,li2020substantial,buitrago2020occurrence,world2003consensus,chowell2004model,eichner2003transmission}. 
We then calculate the corresponding values of $\varepsilon^{+}$ and $\ln\eta$ 
based on the Markovian and non-Markovian approaches. As demonstrated in 
Fig.~\ref{fig:forecasting}f, the positions of the four diseases in the 
($\ln{\eta}$, $\varepsilon^{+}$) plane are consistent with the results of our 
estimations. Because the data were from the reports of laboratory-confirmed 
cases incorporating the effects of the quarantine and distancing from susceptible 
individuals after the confirmation of the diagnosis, $T_{\mathrm{gen}}$ of the 
four diseases are all smaller than the corresponding values of $T_{\mathrm{rem}}$, 
leading to some overestimation for the Markovian forecasting results. 

\subsubsection*{Evaluation of vaccination strategies}

\begin{figure} [ht!]
\centering
\includegraphics[width=\linewidth]{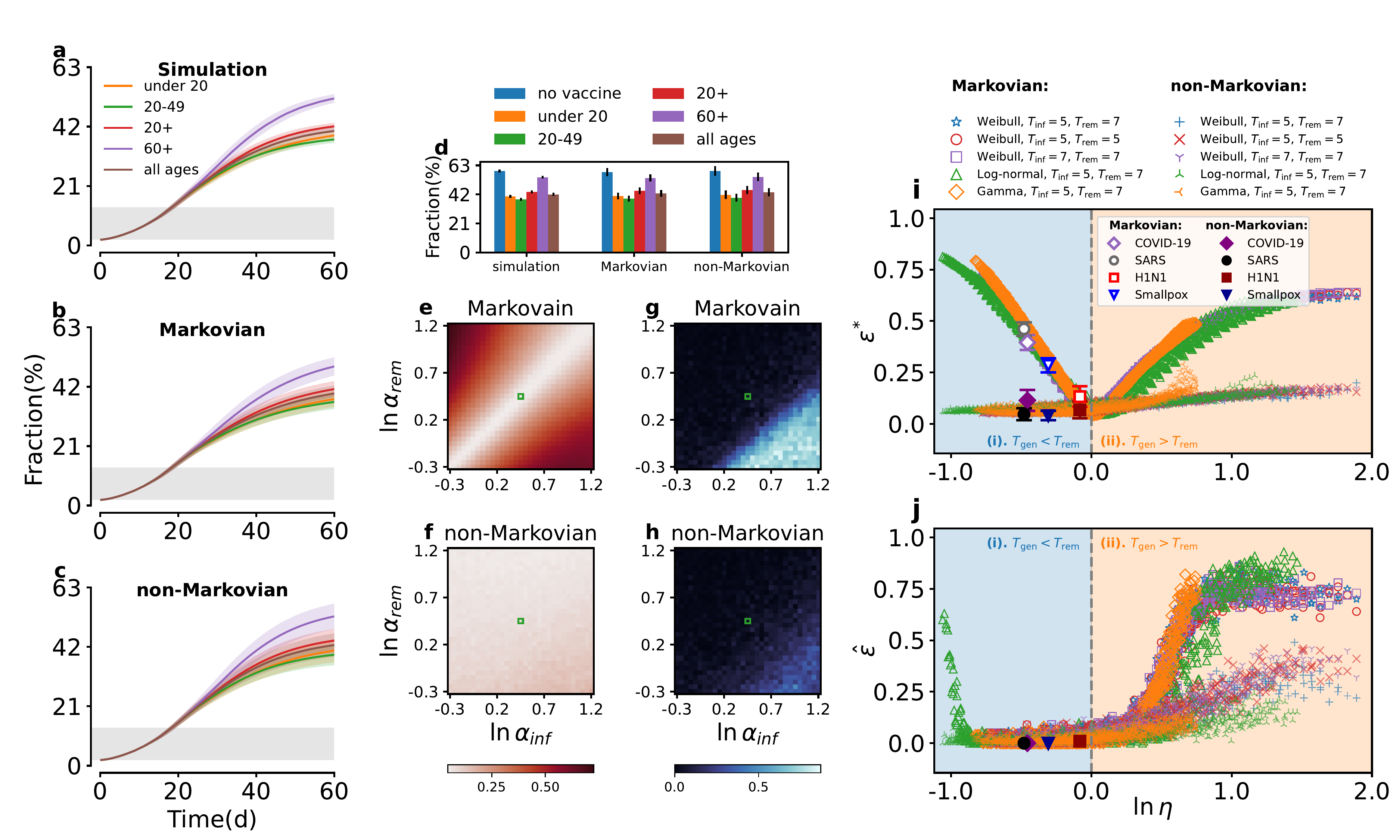}
\caption{\textbf{Evaluation of vaccination strategies}. \textbf{a--c}: For 
simulation parameters chosen according to $\ln{\alpha_{\mathrm{inf}}} = 0.45$ and
$\ln{\alpha_{\mathrm{rem}}} = 0.45$, the cumulative infected fraction (i.e., sum of infected and removed fractions) curves from
Monte Carlo simulations and the corresponding Markovian and non-Markovian theories 
with fitting parameters for five vaccination strategies and the case of no 
vaccination. The average results are obtained from 100 independent realizations
with the shaded regions representing the standard deviations. The gray area marks 
the average cumulative infected fraction for training data selection. 
\textbf{d}: Vector $\delta$ calculated from the results in 
Figs.~\ref{fig:prevention}a--c. \textbf{e--f}: The prevention evaluation errors $\varepsilon^{*}$ of 
Markovian and non-Markovian theories for evaluating the effects of vaccination 
prevention in the parameter plane $(\ln{\alpha_{\mathrm{inf}}},\ln{\alpha_{\mathrm{rem}}})$.
The green squares mark the selected parameters in \textbf{a}--\textbf{c}.
\textbf{g--h}: The optimization failure probabilities $\hat\varepsilon$ arising from the Markovian 
and non-Markovian within the parameter plane $(\ln{\alpha_{\mathrm{inf}}},\ln{\alpha_{\mathrm{rem}}})$.
The green squares mark the selected parameters in \textbf{a}--\textbf{c}.
\textbf{i}: The prevention evaluation errors $\varepsilon^{*}$ from the Markovian and non-Markovian 
theories versus $\ln{\eta}$ under five scenarios of time distribution setting 
for Monte Carlo simulations. The estimated errors for four real diseases 
(COVID-19, SARS, H1N1 and Smallpox) are also shown. 
\textbf{j}: The optimization failure probabilities $\hat\varepsilon$ from the Markovian and non-Markovian 
theories against $\ln{\eta}$ in five different time distribution scenarios 
for Monte Carlo simulations. The optimization failure probabilities for four real diseases 
(COVID-19, SARS, H1N1 and Smallpox) are also presented.}
\label{fig:prevention} 
\end{figure}

In the development and application of a theory for disease spreading, assessing
the effects of different vaccination strategies is an important task. Here we
consider five prioritization strategies for vaccine 
distribution~\cite{bubar2021model}: individuals under 20 years (denoted as $m=1$), 
adults between 20 and 49 years ($m=2$), adults above 20 years ($m=3$), adults 
above 60 years ($m=4$), and all age groups ($m=5$), and implement these strategies 
in Monte Carlo simulations. Fig.~\ref{fig:prevention}a shows the results of 
epidemic evolution in comparison with those without any vaccination intervention ($m=0$), 
where the shape parameters are chosen according to 
$\ln{\alpha_{\mathrm{inf}}}=0.45$ and $\ln{\alpha_{\mathrm{rem}}}=0.45$ 
($T_{\mathrm{gen}} = T_{\mathrm{rem}}$). Figs.~\ref{fig:prevention}b--c show 
the results from the Markovian and non-Markovian theories, respectively, with the 
corresponding fitted parameters for the vaccination strategies in comparison with 
those without vaccination (see {\bf Method} for the detailed procedure of vaccination 
in the theoretical calculation). These results indicate that the Markovian and 
non-Markovian theories yield the correct epidemic evolution and future outbreaks 
under different vaccination scenarios.

To characterize the effectiveness of different vaccination strategies in blocking 
disease transmission, we introduce a vector, $\delta$, whose $m$-th element 
quantifies the cumulative infected fraction with the $m$-th vaccination strategy
in the steady state: $\delta_m = \tilde{c}_m$, for $m=0,\ldots,5$.  
Fig.~\ref{fig:prevention}d shows that the $\delta$ vectors from the Monte Carlo 
simulation, Markovian and non-Markovian theories from 
Figs.~\ref{fig:prevention}a--c, respectively. 

We further introduce a metric, 
the so-called prevention evaluation error $\varepsilon^{*}$, that gauges 
the ability of the Markovian and non-Markovian theories to estimate the total 
effectiveness of vaccination, i.e., measuring the disparity between the results 
calculated by the Markovian or non-Markovian theories 
and those obtained through Monte Carlo simulations 
considering various vaccination strategies (see {\bf Method} for the detailed 
definition of $\varepsilon^{*}$). Figs.~\ref{fig:prevention}e–-f 
show the average values of $\varepsilon^*$ of the two theories in the simulation 
parameter plane using 100 independent realizations, which are similar to those 
in Figs.~\ref{fig:forecasting}d--e, indicating that the error mainly comes from 
the $R_0$ estimation. In general,
the Markovian theory performs well only in the diagonal area of the parameter 
plane where $\alpha_{\mathrm{inf}}=\alpha_{\mathrm{rem}}$, as shown in 
Fig.~\ref{fig:prevention}e, and the non-Markovian theory outperforms the 
Markovian counterpart in most cases, as shown in Fig.~\ref{fig:prevention}f. 

Meanwhile, we assess the ability of both the Markovian and non-Markovian 
theories to detect the optimal vaccination strategy. We also define a quantity, optimization failure probability
$\hat\varepsilon$, to quantify the probability of a theory failing to 
identify the optimal strategy, i.e., that leads to the lowest cumulative infection among the five strategies
(see {\bf Method} for the detailed definition of $\hat\varepsilon$). 
Figs.~\ref{fig:prevention}g–-h illustrate the results of $\hat\varepsilon$ 
for the two theories within the parameter plane $(\ln{\alpha_{\mathrm{inf}}},\ln{\alpha_{\mathrm{rem}}})$.
While the non-Markovian theory still demonstrates superior performance, 
the Markovian approach proves capable of identifying the optimal strategy 
across a significantly larger parameter space compared to 
the Markovian results depicted in Figure~\ref{fig:prevention}e. 

We obtain the relationships between $\varepsilon^*$ and $\ln{\eta}$, 
as well as between $\hat\varepsilon$ and $\ln{\eta}$, as shown
in Fig.~\ref{fig:prevention}i--j with the same five time-distribution scenarios as
in Fig.~\ref{fig:theories}j. In all cases of Fig.~\ref{fig:prevention}i, $\varepsilon^*$ reaches a minimum for
$\ln{\eta} = 0$ and increases as $\ln{\eta}$ deviates from zero. The agreement 
of the results from the five scenarios further illustrates that the forms of 
time distributions or the specific values of $T_{\mathrm{gen}}$, 
$T_{\mathrm{inf}}$, or $T_{\mathrm{rem}}$ play little role in the errors 
in vaccination evaluation. Fig.~\ref{fig:prevention}i also includes the
values of $\varepsilon^*$ for the real-world infectious diseases COVID-19,
SARS, H1N1 influenza, and smallpox, which are consistent with those from the
non-Markovian and Markovian theories. Regarding the results depicted in Fig.~\ref{fig:prevention}j, 
it is observed that the non-Markovian theories consistently outperform the Markovian counterparts.
On the other hand, within a wide range of $\ln\eta$ values around 0, 
the Markovian theories successfully identify the optimal vaccination 
strategy among various commonly employed ones. When the value of $\ln\eta$ significantly deviates from 0, 
Markovian theories become ineffective in determining the optimal strategy.
(Note that on the left side of Fig.~\ref{fig:prevention}j,
we only present the failures of Markovian theories to
identify the optimal strategy in the Monte Carlo simulations with log-normal distribution. 
This is primarily due to the fact that the parameters associated 
with the Weibull and gamma distributions fall outside the acceptable range 
when we keep $T_{\mathrm{inf}}$ and $T_{\mathrm{rem}}$ fixed to modify $\ln\eta$ to a very low value.)
Furthermore, we demonstrate that even when employing Markovian approaches, 
the optimal vaccination strategy can still be determined among the 
five strategies considered for the four distinct real diseases.

Conducting accurate evaluations in prevention serves as the sufficient condition of the successful identification of the optimal strategy. In comparison
to the prevention evaluation errors $\varepsilon^*$ of Markovian theories, the optimization failure probability $\hat\varepsilon$ exhibits a wider
range of $\ln\eta$ values that result in the lowest value. The lack of mathematical continuity among the five strategies is
the primary reason for this. It indicates that there is no smooth transition or mathematical relationship connecting
these strategies, resulting in the rank of the strategies not changing promptly when the value of $\ln\eta$ deviates from 0.
Therefore, only significant errors from the Markovian 
theories can result in the failure to detect the optimal strategy.
Based on this analysis, the extent to which $\ln\eta$ deviates from 0, 
leading to the failure of Markovian theories, as well as whether such failure will occur,
depends on the selection of the tested strategies.
            
\section*{Discussion}

The COVID-19 pandemic has emphasized the importance of investigating disease 
transmission in human society through modeling. Empirical observations have 
consistently demonstrated strong memory effects in real-world transmission 
phenomena. The initial transient stage of an epidemic is critical for data 
collection, prediction, and articulation of control strategies, but an accurate
non-Markovian model presents difficulties. In contrast, a Markovian model offers 
great advantages in parameter estimation, computation, and analyses. Uncovering 
the conditions under which Markovian modeling is suitable for transient epidemic 
dynamics is necessary.   

We have developed a comprehensive mathematical framework for both Markovian and 
non-Markovian compartmentalized SIR disease transmissions in an age-stratified 
population, which allows us to identify two types of equivalence between Markovian
and non-Markovian dynamics: in the steady state and transient phase of the 
epidemic. Our theoretical analysis reveals that, in the steady state, 
non-Markovian (memory-dependent) transmissions are always equivalent to the 
Markovian (memoryless) dynamics. However, transient-state equivalence is 
approximate and holds when the average generation and removal times match each 
other. In particular, when the average generation time is approximately equal to
the average removal time, the disease transmission and removal of an infected 
individual exhibit a memoryless correlation, thereby minimizing the memory effects
of the dynamical process. This results in highly accurate results from the 
Markovian theory that captures the characteristics of memory-dependent 
transmission based solely on the early epidemic curves. Our analysis also suggests
that the Markovian accuracy is mainly determined by the value of 
generation-to-removal time ratio in disease transmission, where a larger-than-one 
(smaller-than-one) ratio can lead to underestimation (overestimation) of the 
basic reproduction number and epidemic forecasting, as well as the errors in 
the evaluation of control or prevention measures. The estimation accuracy 
primarily depends on this ratio, but is not significantly affected by the specific 
values and distribution forms of the various times associated with the epidemic. This property exhibits substantial practical importance, because the average generation and removal times can be readily assessed
based on sparse data collected from the transient phase of the epidemic, but 
to estimate their distributions with only sparse data is infeasible. These 
results provide deeper quantitative insights into the influence of memory effects 
on epidemic transmissions, leading to a better understanding of the connection and
interplay between Markovian and non-Markovian dynamics. 

There were previous studies of the equivalence between Markovian and non-Markovian
transmission in the SIS model~\cite{starnini2017equivalence,feng2019equivalence,cator2013susceptible}. However, these studies addressed the steady-state equivalence 
rather than the transient-state equivalence. To our knowledge, our work is the 
first to investigate the transient-state equivalence of the SIR model. In
addition, previous studies mainly examined the impact of the average generation 
time on the transmission dynamics, such as how the shape of the generation time 
distribution affects the estimation of $R_0$~\cite{wallinga2007generation} or the 
use of serial time distributions in estimating $R_0$ during an 
epidemic~\cite{park2021forward}. There was a gap in the literature regarding how 
generation times affect the accuracy of different models. Our paper fills this gap
by providing a criterion for using Markovian frameworks to model memory-dependent 
transmission based on the relationship between the average generation and removal 
times.

From an application perspective, our study suggests that the impact of the time 
distribution forms on Markovian estimation accuracy is minimal, making it easier 
to select models between Markovian and non-Markovian dynamics in the initial 
outbreak of an epidemic based only on the generation-to-removal time ratio. This 
insight is especially useful since detailed time distribution forms are often 
harder to detect than their corresponding mean values. In addition, we note that
in previous studies, it was observed that in various scenarios,
serial intervals, albeit with larger variances, 
are anticipated to possess a consistent mean value with the average generation 
time and are more straightforward to measure~\cite{ferretti2020quantifying,park2021forward,svensson2007note,klinkenberg2011correlation,te2013estimating, champredon2018equivalence}. Given the practical difficulties in observing the generation time, our 
finding of minimal impact from the distribution forms suggests that the average 
serial interval can be utilized as a substitute of the average generation time to 
determine the applicability of the Markovian theories for modeling purposes 
without compromising accuracy, although numerous studies have indicated that 
replacing the generation time distribution with the serial interval distribution 
may affect the analysis of transmission dynamics~\cite{park2021forward,britton2019estimation,ganyani2020estimating}.
Meanwhile, based on the Eq.~\eqref{eq:r0_estimation}, if we determine the ratio of generation-to-removal time, 
the estimated $R_0$ obtained through the Markovian approach can be adjusted to approximate the true.
And our web-based application showcases the demonstration of rectifying $R_0$ and epidemic forecasting \cite{Validity_MM}.

Our study highlights the critical importance of accurately quantifying $R_0$ for 
achieving precise epidemic forecasting and prevention evaluation. A previous 
work~\cite{dietz1993estimation} revealed that the value of $R_0$ depends on 
three key components: the duration of the infectious period (e.g., 
$\psi_{\mathrm{rem}}(\tau)$), the probability of infection resulting from a 
single contact between an infected individual and a susceptible one (e.g., 
$\psi_{\mathrm{inf}}(\tau)$), and the number of new susceptible individuals 
contacted per unit of time. However, given the practical limitations inherent 
in obtaining all three components, numerous methods have been developed for 
estimating $R_0$. Although our work presents a specific approach, which fits the 
parameters of exponential or non-exponential time distribution by using the 
initial outbreak curves, it is not the only one available. For example, when 
contact patterns are unknown, $R_0$ can be estimated by fitting the growth rate 
$g$ and the generation time distribution $\psi_{\mathrm{gen}}(\tau)$, and then 
applying them in the Euler-Lotka equation~\cite{dietz1993estimation,wallinga2007generation,park2021forward}. However, since the focus of our work is on epidemic 
forecasting and evaluation of prevention measure, $R_0$ can be directly calculated
once $\psi_{\mathrm{inf}}(\tau)$ and $\psi_{\mathrm{rem}}(\tau)$ are fitted, 
without requiring the fitting of any additional quantity. The estimation of $R_0$ 
can also be achieved by using data in the steady state, such as the final size of 
an epidemic or equilibrium conditions~\cite{dietz1993estimation}. However, this 
method is not suitable for the transient phase where only early-stage curves are 
available. Utilizing the approach delineated in this paper is practically more
appropriate for estimating $R_0$.

While our study focused on transmission within the SIR framework, extension to 
SEIR or SIS models is feasible. While we emphasized the significance of the 
transient-state equivalence in disease transmission, transient dynamics are 
more relevant or even more crucial than the steady state in nonlinear dynamical
systems~\cite{LT:book}. For example, in ecological systems, transient dynamics 
play a vital role in empirical observations and are therefore a key force driving 
natural evolution~\cite{hastings1994persistence,hastings2004transients,hastings2016timescales,hastings2018transient,MACFGHLPSZ:2020,MLG:2020}. In neural dynamics, 
transient changes in neural activity can mediate synaptic plasticity, a crucial 
mechanism for learning and memory~\cite{sjostrom2001rate,martin2000synaptic,kandel2001molecular}. Therefore, the identification of suitable conditions for choosing 
between Markovian and non-Markovian dynamics may not be limited to transmission 
dynamics alone and may serve as a valuable reference for other fields as well.

Taken together, our study establishes an approximate equivalence between Markovian
and non-Markovian dynamics in the transient state, assuming that time 
distributions follow Weibull forms (see Supplementary Note 3 for details). While the applicability of our findings to 
most synthetic and empirical distributions has been analyzed qualitatively, a 
quantitative analysis requires further studies. For extreme cases with non-Weibull
distributions, the transmission should be evaluated using other specific methods. 
While we have provided a qualitative analysis of the mechanism underlying why time
distribution forms have minimal impacts on the errors of Markovian estimations, a 
more rigorous theoretical analysis is needed and requires further exploration. In 
addition, due to the complexity of the nonlinear transmission, our study has 
produced a semi-empirical relationship to estimate the overestimation and 
underestimation of Markovian methods. Further research is required to develop a 
rigorous formula that can accurately predict these effects.

\section*{Methods}

\subsection*{Monte Carlo simulation}
    
In the simulation, we classify $N$ individuals into $n$ subgroups based on the 
age distribution $\mathbf{\textit{p}}$. The index of the subgroup to which an individual 
belongs is denoted by $l$ (where $0 \le l \le n-1$), and the index of the 
individual within the subgroup is denoted by $u$ (where $0 \le u \le \mathbf{\mathit{p}}_lN$). 
The state of the $u$-th individual in the $l$-th subgroup is represented by 
$X_{lu}$, which includes the states $S$ (susceptible), $I$ (infected), $W$ 
(recovered), and $D$ (dead), where $W$ and $D$ both represent $R$ (removed). For 
each individual, we also record the absolute time of infection and removal using 
two variables: $T_{lu}^{\mathrm{inf}}$ and $T_{lu}^{\mathrm{rem}}$, respectively. 
The absolute time of the system is denoted by $T$, and we implement the total 
spreading simulation step by step using a finite time step $\Delta T$ as follows:
\begin{itemize}
\item [i)] 
Initialization: set $T=0$, $X_{lu}$ for every individual is set to $S$.
\item [ii)] \vspace*{-0.1in}
Set infection seeds: choose a set of individuals as the infection seeds and the 
corresponding $X_{lu}$ are set to $I$, the corresponding $T_{lu}^{\mathrm{inf}}$ 
are set to 0, and $T_{lu}^{\mathrm{rem}}$ are set to a random value following the 
removal time distribution $\psi_{\mathrm{rem}}(\tau)$.
\item[iii)] \vspace*{-0.1in}
Infection of one step: calculate the infection rate, 
$\hat{\omega}_{lu}^{\mathrm{inf}}(T)$, of infected individual $u$ in age group 
$l$ during the current time step by 
\begin{align} \nonumber
\hat{\omega}_{lu}^{\mathrm{inf}}(T) = 1 - \Psi_{\mathrm{inf}}(T - T_{lu}^{\mathrm{inf}} + \Delta T) / \Psi_{\mathrm{inf}}(T - T_{lu}^{\mathrm{inf}}). 
\end{align}
The probability $\bar{\omega}_{l}^{\mathrm{inf}}(T)$ of each susceptible 
individual in age group $l$ being infected can be calculated by 
\begin{align} \nonumber
\bar{\omega}_{l}^{\mathrm{inf}}(T) = \sum_{m = 1}^{n}{\mathbf{\mathit{p}}_m[1 - (1 - \frac{\sum_{v\in \mathcal{I}_m}{\hat{\omega}_{mv}^{\mathrm{inf}}(T)}}{\mathbf{\mathit{p}}_mN})^{k\mathbf{\mathit{A}}_{lm}}]},
\end{align}
where $\mathcal{I}_m$ is the index set of the infected individual in age group 
$m$. The number of the susceptible individuals being infected in a age group 
follows a binomial distribution $B(s_l(T)\mathbf{\mathit{p}}_lN,\bar{\omega}_{l}^{\mathrm{inf}}(T))$,
where $s_l(T)$ denotes the fraction of susceptible individuals in age group $l$ 
at time $T$. Then generate a random number $N_l(T)$ following this binomial 
distribution and set the $N_l(T)$ individuals as $I$ state. The 
corresponding $T_{lu}^{\mathrm{inf}}$ of the new infected individuals are set 
to the current $T$ and $T_{lu}^{\mathrm{rem}}$ are set to $T_{\mathrm{rem}} + T$, 
where the random $T_{\mathrm{rem}}$ follow the removal time distribution 
$\psi_{\mathrm{rem}}(\tau)$.
\item[iv)] \vspace*{-0.1in}
Check if $T_{lu}^{\mathrm{rem}}$ of each infected individual is during the 
current time step. If this condition is satisfied, set their state to $D$ with 
a probability of death and to $W$ otherwise. Then let $T = T + \Delta T$.
\item[v)] \vspace*{-0.1in}
Repeat the process iii) and iv), until no individual with $I$ index exists.
\end{itemize}

\subsection*{Time distributions}
In the Monte Carlo simulations, we employ three types of time distributions, 
i.e., Weibull, log-normal, and gamma, to describe the memory-dependent transmission process.

For Weibull time distribution, it follows:
\begin{align} \nonumber
\psi(\tau) = \frac{\alpha}{\beta}(\frac{\tau}{\beta})^{\alpha - 1}e^{-(\frac{\tau}{\beta})^{\alpha}},
\end{align}
where $\alpha$ and $\beta$ denote the shape and scale parameters, respectively.

The log-normal time distribution is defined as follows:
\begin{align} \nonumber
\psi(\tau) = \frac{1}{\tau\beta\sqrt{2\pi}}\exp(-\frac{(\ln\tau - \alpha)^2}{2\beta^2}).
\end{align}

The gamma time distribution is expressed as follows:
\begin{align} \nonumber
\psi(\tau) = \frac{1}{\Gamma(\alpha)\beta^\alpha}\tau^{\alpha-1}e^{-\frac{\tau}{\beta}},
\end{align}
where $\Gamma(\cdot)$ denotes gamma function, while $\alpha$ and $\beta$ represent the shape and scale parameters, respectively.

For each type of time distribution, denoted as $\psi(\tau)$, 
it can be either $\psi_{\mathrm{inf}}(\tau)$ or $\psi_{\mathrm{rem}}(\tau)$. 
Similarly, the parameter $\alpha$ can take either $\alpha_{\mathrm{inf}}$ or $\alpha_{\mathrm{rem}}$, 
and the parameter $\beta$ can be either $\beta_{\mathrm{inf}}$ or $\beta_{\mathrm{rem}}$.

Additionally, the survival function could calculated by:
\begin{align}\nonumber
    \Psi(\tau) = \int_{\tau}^{+\infty}{\psi(\tau')d\tau'},
\end{align}
while the hazard function could deducted by:
\begin{align}\nonumber
    \omega(\tau) = \frac{\psi(\tau)}{\int_{\tau}^{+\infty}{\psi(\tau')d\tau'}}.
\end{align}
The survival function $\Psi(\tau)$ can be either $\Psi_{\mathrm{inf}}(\tau)$ or $\Psi_{\mathrm{rem}}(\tau)$, 
and the hazard function $\omega(\tau)$ can take either $\omega_{\mathrm{inf}}(\tau)$ or $\omega_{\mathrm{rem}}(\tau)$.

\subsection*{Derivation of semi-empirical estimation of basic reproduction number}

Intuitively, the period of $T_{\mathrm{rem}}$ can accommodate $T_{\mathrm{rem}}/T_{\mathrm{gen}}=1/\eta$ time intervals of length $T_{\mathrm{gen}}$, corresponding to the result of $1/\eta$ generations of infections. This can lead to an exponential increase in the number of infections during $T_{\mathrm{rem}}$. This intuition suggests a relationship between the fitted basic reproduction number $\hat{R}_0$ and the actual $R_0$, which can be expressed as an exponential function: 
\begin{align}\nonumber
    \hat{R}_0 = R_0^{f(1/\eta)},
\end{align}
where $f(\cdot)$ is a monotonically increasing function that satisfies three conditions. First, $f(1) = 1$, indicating that $\hat{R}_0$ can be accurately estimated when $T_{\mathrm{gen}} = T_{\mathrm{rem}}$. Second, $f(0) = 0$, meaning that if $T_{\mathrm{rem}}$ is an extremely small fraction of $T_{\mathrm{gen}}$, the transmission will take a long time to reach the steady state, causing the curve to be flat in the initial stage and potentially causing the Markovian fitting to produce the estimate $\hat{R}_0=1$. Third, $f(+\infty) = +\infty$, implying that if $T_{\mathrm{rem}}$ is extremely large compared to $T_{\mathrm{gen}}$, the transmission will quickly reach the final prevalence, causing the Markovian fitting to give an extremely large estimate of $\hat{R}_0$. Because the actual transmission process involves many complicated nonlinear relationships, identifying the specific form of the function $f(\cdot)$ is a challenging task. We thus assume
\begin{align} \nonumber
     f(x) = x^a, 
\end{align}
where $a$ is an unknown positive coefficient. This leads to Eq.~\eqref{eq:r0_estimation}.

\subsection*{Definition of errors}

The difference $\varepsilon$ between non-Markovian and the corresponding Markovian results calculated from
Eqs.~(\ref{eq:inferred_gamma}--\ref{eq:inferred_mu}) is defined as: 
\begin{align} \nonumber
\varepsilon = \sum_{x, x^*\in\{(s, s^*),(i, i^*),(r, r^*)\}}{\frac{\Vert x^* - x\Vert_{2, T}}{\Vert x\Vert_{2, T}}}
\end{align}
where the pairs $(s, s^*)$, $(i,i^*)$ and $(r,r^*)$ correspond to the 
non-Markovian and Markovian susceptible, infected and removed curves, 
respectively. The 2-norm $\Vert\cdot\Vert_{2, T}$ on time duration $T$ ensures 
that $\varepsilon$ measures the ``distance'' between non-Markovian and the 
Markovian transient states. It is not appropriate to set $T$ as the total transmission 
period because the cumulative infected fraction approaches the final value asymptotically, 
making it difficult to determine the exact time point of the steady state. 
To address this issue, we choose $T$ as $[0, \tilde{t}_\theta]$, where $\tilde{t}_\theta$ is 
the time when the cumulative infected fraction $c(t)$ reaches the $\theta$ percentile 
point within the range that spans from its initial value ($\acute{c}$) to its final value ($\tilde{c}$).
The value of $\theta$ in Fig.~\ref{fig:theories}j is selected as 50 
(see Supplementary Note 6 for more selection of $\theta$ and detailed analysis).

The forecasting error $\varepsilon^{+}$ that 
evaluates whether a theory overestimates or underestimates the steady-state cumulative
infected fraction is defined as:
\begin{align} \nonumber
\varepsilon^{+} = \frac{c(\tilde{t}) - \hat{c}(\tilde{t})}{\hat{c}(\tilde{t})},
\end{align}
where $\tilde{t}$ denotes the time when the stochastic simulation reaches the 
steady state when no infection occurs in the population, $c(\tilde{t})$ and 
$\hat{c}(\tilde{t})$ are the cumulative infected fractions from theory and 
simulation, respectively. A positive value of $\varepsilon^{+}$ indicates overestimation, 
whereas a negative value indicates underestimation.

The prevention evaluation error $\varepsilon^{*}$, that gauges 
the ability of the Markovian and non-Markovian theories to estimate the total 
effectiveness of vaccination, is defined as:
\begin{align} \nonumber
\varepsilon^* = \frac{\Vert\hat{\delta} - \delta\Vert_2}{\Vert\hat{\delta}\Vert_2},
\end{align}
where $\hat{\delta}$ is the result from Monte Carlo simulation and
$\Vert \cdot \Vert_2$ is the 2-norm of a vector.

The optimization failure probability $\hat\varepsilon$, which measures the
probability that a theory fails to identify the optimal vaccination strategy,
is defined as:
\begin{align} \nonumber
\hat{\varepsilon} = \frac{\sum_{l = 1}^{z}{\xi^{(l)}}}{z},
\end{align}
where
\begin{align} \nonumber
\xi^{(l)} =
    \begin{cases}
        0 & \text{if } \argmin\hat{\sigma}^{(l)} = \argmin\sigma^{(l)} \\
        1 & \text{otherwise}
    \end{cases},
\end{align}
$\hat{\sigma}^{(l)}$ and $\sigma^{(l)}$ represent the vectors, $\hat{\sigma}$ and $\sigma$, 
for the $l$-th experiment, respectively, and $z$ denotes the total number of experiments (in this paper, $z$ is set to 100). 
Consequently, $\hat{\varepsilon}$ quantifies the fraction of experiments in which a theory fails to identify 
the optimal vaccination strategy, and serves as a measure of the probability of failure in optimizing the vaccination strategy.

\subsection*{Fitting method}
    
Because the removal process is independent of the infection one, we divide the 
fitting method into two parts: removal parameter fitting and infection parameter 
fitting. Specifically, we use $c_{\mathrm{init}}^*$ and $r_{\mathrm{init}}^*$ to 
denote the cumulative infected and removed fractions in the initial stage of a 
Monte Carlo simulation. These two types of data are substituted into the 
Eq.~\eqref{eq:r_t} to fit the parameters of $\psi_{\mathrm{rem}}(\tau)$. Likewise,
we use $c_{l, \mathrm{init}}^*$ to denotes the cumulative infected fraction of age
group $l$ in the initial stage of a Monte Carlo simulation and 
$s_{l, \mathrm{init}}^* = 1 - c_{l, \mathrm{init}}^*$ represents the corresponding
susceptible fraction. After obtaining the removal parameters, these two types of
data are put into Eq.~\eqref{eq:s_t} to fit the infection time distribution 
parameters.
    
In our study, we selected the curves of all states that occurred prior to the time
point at which the cumulative infected fraction reached a specific percentile 
(e.g., $20\%$) situated between the initial and steady cumulative infected 
fractions, as the training data. Choosing a specific time period as the training 
data may not be appropriate, as it can result in an overabundance of data points 
for fitting due to some instances of fast transmission already having reached the 
steady state, while some instances of slow transmission may not have spread out 
yet, leading to too few data points.
    
\subsection*{Vaccination method}
   
We assume that the individuals will build enough immune protection from the 
disease $\kappa$ days after vaccination with the probability $\rho$. In Monte 
Carlo simulations, the susceptible individual who gets vaccinated and the 
corresponding time $T_{\mathrm{vac}}$ are associated with the probability $\rho$,
and the absolute time becomes $T_{\mathrm{vac}} + \kappa$. If this individual 
has not been infected, he/she will be set to a state called protected state,
indicating that this individual is protected from the disease.

When a fraction of individuals in age group $l$ get vaccinated, the detailed 
vaccination fraction $\upsilon_l$, susceptible time $t_{\mathrm{vac}}$ and the 
fraction of susceptible individuals $s_l(t_{\mathrm{vac}})$ are recorded. When
the absolute time reaches $T_{\mathrm{vac}} + \kappa$, the corresponding value 
of $s_l(t_{\mathrm{vac}} + \kappa)$ will be set as  
$s_l(t_{\mathrm{vac}} + \kappa)\rho\frac{\upsilon_l}{s_l(t_{\mathrm{vac}})}$.

\section*{Data Availability}

\vspace*{-0.2in}
All relevant data are available at {\color{blue}\url{https://github.com/fengmi9312/Validity-of-Markovian-for-Memory/tree/main/FigureData}}.

\vspace*{-0.2in}
\section*{Code Availability}

\vspace*{-0.2in}
The web-based application can be visited at {\color{blue} \url{https://cns.hkbu.edu.hk/toolbox/Validity-of-Markovian-for-Memory/main.html}}. The GitHub repository, which includes the source code for all the figure results, the web-based application, and an additional Python application, can be accessed at
{\color{blue} \url{https://github.com/fengmi9312/Validity-of-Markovian-for-Memory.git}}.

\vspace*{-0.2in}
\section*{Acknowledgments}

\vspace*{-0.2in}
This work was supported by the Hong Kong Baptist University (HKBU) Strategic Development Fund. 
This research was conducted using the resources of the High-Performance Computing Cluster Centre at HKBU, 
which receives funding from the Hong Kong Research Grant Council and the HKBU.
Y.-C.L was supported by the Office of Naval Research through Grant
No.~N00014-21-1-2323.

\vspace*{-0.2in}
\section*{Author Contributions}
\vspace*{-0.2in}
M.F., L.T. and C.-S.Z. designed research; M.F. performed research; L.T. and C.-S.Z. contributed analytic tools; M.F., L.T. and C.-S.Z. analysed data; M.F., L.T., Y.-C.L. and C.-S.Z. discussed the results and wrote the paper.

\vspace*{-0.2in}
\section*{Competing Interests}

\vspace*{-0.2in}
The authors declare no competing interests.

\vspace*{-0.2in}
\section*{Correspondence}

\vspace*{-0.2in}
To whom correspondence should be addressed: liangtian@hkbu.edu.hk, cszhou@hkbu.edu.hk

\bibliographystyle{naturemag}

\end{document}